\documentclass[fleqn,usenatbib]{mnras}

\usepackage{newtxtext,newtxmath}
\usepackage{caption}
\usepackage{subcaption}

\usepackage[T1]{fontenc}

\DeclareRobustCommand{\VAN}[3]{#2}
\let\VANthebibliography\thebibliography
\def\thebibliography{\DeclareRobustCommand{\VAN}[3]{##3}\VANthebibliography}

\usepackage{graphicx}	
\usepackage{amsmath}	

\newcommand{\OII}{[\hbox{{\rm O}\kern 0.1em{\sc ii}}]}

\newcommand{\sigmagas}{$\sigma_{g}$}

\newcommand{\kms}{km~s$^{-1}$}

\newcommand{\halpha}{H$\alpha$}
\newcommand{\Halpha}{H$\alpha$}
\newcommand{\Lyalpha}{Ly$\alpha$}
\newcommand{\hbeta}{H$\beta$}
\newcommand{\OIII}{[\hbox{{\rm O}\kern 0.1em{\sc iii}}]}
\newcommand{\NII}{[\hbox{{\rm N}\kern 0.1em{\sc ii}}]}
\newcommand{\SII}{[\hbox{{\rm S}\kern 0.1em{\sc ii}}] }
\newcommand{\OIIIHbeta}{$\log([\hbox{{\rm O}\kern 0.1em{\sc iii}}]/H\beta)$}
\newcommand{\NIIHalpha}{$\log([\hbox{{\rm N}\kern 0.1em{\sc ii}}]/H\alpha)$}
\newcommand{\surfacebrightnesslimit}{erg cm$^{-2}$ s$^{-1}$ arcsec$^{-2}$}
\newcommand{\fluxunits}{erg s$^{-1}$ cm$^{-2}$ }

\graphicspath{{./}{figures/}}

\title[Abell 2390 Brightest Cluster Galaxy]{Extended Line Emission in the BCG of Abell 2390}

\author[Alcorn et al.]{
Leo Y. Alcorn,$^{1}$\thanks{E-mail: ly.alcorn@utoronto.ca}
H.K.C Yee,$^{1}$
Laurent Drissen, $^{2,3}$
Carter Rhea, $^{3,4}$
Suresh Sivanandam, $^{1,5}$
\newauthor Julie Hlavacek-Larrondo, $^{3,4}$
Bau-Ching Hsieh, $^{6}$
Lihwai Lin, $^{6}$
Yen-Ting Lin, $^{6}$
Qing Liu, $^{1,5}$
Adam Muzzin, $^{7}$
\newauthor Allison Noble $^{8}$
and Irene Pintos-Castro$^{9}$
\\
$^{1}$David A. Dunlap Department of Astronomy \& Astrophysics, University of Toronto, 50 St. George St., Toronto, ON M5S 3H4, Canada\\
$^{2}$Département de Physique, de génie physique et d'optique, Université Laval, Québec, QC, G1V 0A6, Canada\\
$^{3}$Centre de Recherche en Astrophysique du Québec (CRAQ)\\
$^{4}$Département de Physique, Université de Montréal, Succ. Centre-Ville, Montréal, Québec, H3C 3J7, Canada\\
$^{5}$Dunlap Institute of Astronomy and Astrophysics, University of Toronto, 50 St. George St, Toronto, ON, Canada\\
$^{6}$Institute of Astronomy and Astrophysics, Academia Sinica, No. 1, Section 4, Roosevelt Road, Taipei 10617, Taiwan\\
$^{7}$Department of Physics and Astronomy, York University, 4700 Keele Street, Toronto, Ontario, ON MJ3 1P3, Canada\\
$^{8}$School of Earth and Space Exploration, Arizona State University, Tempe, AZ, 85287, USA\\
$^{9}$Centro de Estudios de Física del Cosmos de Aragón, Pl. San Juan, 1, 44001 Teruel, Spain
}

\date{Accepted XXX. Received YYY; in original form ZZZ}

\pubyear{2022}

\begin{document}
\label{firstpage}
\pagerange{\pageref{firstpage}--\pageref{lastpage}}
\maketitle

\begin{abstract}
We report CFHT/SITELLE imaging Fourier Transform Spectrograph observations of the Brightest Cluster Galaxy (BCG) of galaxy cluster Abell 2390 at $z=0.228$. The BCG displays a prominent cone of emission in \Halpha, \hbeta, \NII, and \OII\ to the North-West with PA = 42$^o$, 4.4\arcsec\ in length (15.9 kpc), which is associated with elongated and asymmetric Chandra soft X-ray emission. The \Halpha\ flux map also contains a “hook” of \Halpha\ and \NII\ emission resulting in a broadened northern edge to the cone. Using SITELLE/LUCI software we extract emission line flux, velocity, velocity dispersion, and continuum maps, and utilize them to derive flux ratio maps to determine ionization mechanisms and dynamical information in the BCG's emission line region. The Baldwin-Phillips-Terlevich diagnostics on the BCG cone indicate a composite ionization origin of photoionization due to star formation and shock. Strong LINER-like emission is seen in the nuclear region which hosts an AGN. As Abell 2390 is a cool-core cluster, we suggest that the cooling flow is falling onto the central BCG and interacting with the central AGN. The AGN produces jets that inflate “bubbles” of plasma in the ICM, as is often observed in local galaxy clusters. Furthermore, combining signs of AGN activities from radio, optical emission line and X-ray data over a large range of physical scale, we find evidence for three possible episodes of AGN activity in different epochs associated with the Abell 2390 BCG.
\end{abstract}




\section{Introduction} \label{sec:intro}

The Brightest Cluster Galaxies (BCGs) of galaxy clusters are a unique population of galaxies, with marked differences from typical cluster galaxies \citep[e.g.][]{lin_new_2010}.
They are the brightest and most massive galaxies observed in our universe, and are located near the galaxy cluster central potential \citep{Oegerle1991,Lin2004,Lauer2014}.
BCG dark matter and stellar halos transition smoothly into the dark matter halo of the entire cluster, and into the diffuse intra-cluster light \citep[see][and references therein]{montes_faint_2022}.
The majority of BCGs are slow-rotating elliptical galaxies \citep{Fisher1995,Tran2008,Brough2011,Jimmy2013}, but in contrast to typical slow-rotating ellipticals in clusters, they are more likely to have strong optical nebular emission  \citep{McNamara1993,Crawford1999, Fogarty2015, Hamer2016} in certain environmental conditions such as cool-core clusters \citep{Heckman1989,Cavagnolo2008,McDonald2010,Tremblay2015,Hogan2017,Calzadilla2022}.

In cool-core clusters, or, clusters which have a cooling flow, the density of the density of the intra-cluster medium (ICM) is inversely proportional to the cooling time, so cooling flows occur when the radiative cooling time of the cluster gas is significantly shorter than the age of the cluster.
The density of the ICM at the centers of clusters increases because cool ICM gas is compressed by the ICM outside the cluster center.
In order to maintain the hydrostatic equilibrium of the cluster, hot ICM gas begins to flow inward, toward the peak of the cluster dark matter halo and the BCG.
This flow of hot gas in toward the center of the cluster is the "cooling flow" in cool-core clusters \citep{Fabian1994, Hudson2010, Blanton2010}.

The BCGs of cool-core clusters are associated with peaks in X-ray emission and the previously mentioned optical nebular emission \citep{Crawford1999,McNamara2007,Cavagnolo2010,Hudson2010, Fabian2012}.
They are also often hosts of radio-loud active galactic nuclei (AGN), which create radio lobes and jets \citep{McNamara2000,Egami2006, Lin2007,Ea2010a}.
When the in-falling gas interacts with the supermassive black hole (SMBH) in a BCG, this can transition the SMBH to its active (kinetic) phase \citep{Fabian2012, Hamer2016}, although in contrast accretion can also result in a radiative phase \citep{Churazov2005,Russell2013,Hlavacek-Larrondo2013}.
The jet activity of a SMBH (AGN feedback) can force pristine gas out of the galaxy and quench star formation (negative feedback), or compress the gas and cause a starburst \citep{McNamara2006,Ea2010a}.
The AGN of cool-core BCGs have been observed to inflate X-ray cavities from rising buoyant bubbles of plasma fuelled by the jets, which entrain cool gas from the galaxy behind them \citep[e.g.][]{Pope2010,David2011,Fabian2016,McNamara2016,Su2017,Tremblay2018,Duan2018,Russell2019,Chen2019b,Smith2021,Zhang2022}, drawing them outward from the BCG, creating X-ray profiles which deviate from circular symmetry of the X-ray surface brightness profile and optical emission line nebulae that extend out from the BCG.
Bubbles and entrained gas can be disrupted via cooling instabilities, merger activity, and gas sloshing, which create bent or horseshoe-shaped emission-line nebulae observed in \halpha \citep{Fabian2003,Ueda2018}.
Examples of this process in the local universe can be seen in NGC1275 of the Perseus cluster \citep{Churazov2000,Fabian2003,Aharonia2018}, and NGC4696 of the Centaurus cluster \citep{Fabian2005}, among many others.

Abell 2390 (hereafter, A2390) is a massive cluster at $z=0.228$, with $M_{200}\sim 1.3\times 10^{15} M_{\odot}$ and $\sigma_{v,cluster}\sim 1100$ $km/s$ \citep{carlberg_galaxy_1996}.
It hosts a BCG with a prominent extended nebular emission line region and an asymmetric X-ray profile.
A2390\footnote{The cluster center is located at a Right Ascension of 21:53:36 and a Declination of +17:41:43 in the J2000 epoch \citep{Haines2013}} was observed as part of the CNOC1 cluster survey on CFHT using the MOS multi-object spectrograph \citep{Yee1996a,Yee1996, Abraham1996}, providing a catalog of redshifts, photometry, spectral feature measurements and morphological indices for 371 cluster galaxies.
The cluster contains several prominent strong gravitational lensing arcs, making it a popular target for high-redshift astronomy in addition to previously discussed dense evolved cluster studies \citep[e.g.][]{Pierre1996,Bezecourt1997,Frye1998,Pello1999,Balogh2000,Li2009,Stroe2017}.
X-ray data of A2390 indicate the cluster has a cool core and anisotropic X-ray morphology \citep{Allen2001, Hlavacek-Larrondo2011, Sonkamble2015}, and it is a dynamically relaxed, non-merging cluster with a strong radio source \citep{Abraham1996, Augusto2006a, Egami2006}.

The BCG of A2390 (named J21536+1741 in \citet{Smail2002} and radio source center located at Right Ascension 21:53:36.82670 Declination +17:41:43.7260 in \citet{Patnaik1992}) was observed in multi-wavelength studies \citep[e.g.][]{Hutchings2000, Egami2006, Augusto2006a}.
The BCG contains several interesting and unusual properties -- an active nucleus with a flat spectral index compact radio source in the core, and a strong \halpha, \hbeta, \NII, and \Lyalpha\ emission line extended region with position angle 42.7$^{o}$ northwest of the nuclear region, referred to as a "cone".
This interplay between the BCG, the SMBH, the surrounding hot ICM, and the cooling gas make this galaxy and its cone an ideal laboratory for studying the extreme physics of AGN jets, the most massive galaxies in the universe, and their effect on the cluster environment, as well as the effect of cold gas accretion onto a central galaxy AGN from the cluster's cooling gas flow.

In this paper we present nebular emission line spectral imaging of the central cD-type BCG of A2390.
The data are obtained in three band-limiting filters designed for the detection of $z\sim0.25$ diagnostic optical nebular emission lines (e.g. \halpha, \NII, \hbeta, \OIII, \OII).
This project is part of an ongoing $z\sim0.25$ cluster survey reported in \citet{Liu2021} using SITELLE on the Canada-France-Hawaii Telescope (CFHT), a Fourier Transform Integral Field Spectrograph \citep{Drissen2019}.
The $11'\times11'$ SITELLE field of view makes it ideal for studying galaxy clusters at low to intermediate redshifts, which can extend out to tens of arcminutes on the sky.
By combining the imaging emission-line data from SITELLE with multi-wavelength data from radio to X-ray, we present a possible AGN activity scenario that connects these data.

This paper is structured as follows:
Section \ref{sec:sitelleobs} discusses our CFHT/SITELLE data and the processing of these data cubes.
Observing conditions and data are reported in Section \ref{sec:cfhtsitelle}, internal data processing and galaxy cataloguing are the subjects of Section \ref{sec:elgfinding}, and the emission line characterization software LUCI and its use in this project is introduced in Section \ref{sec:lucifits}.
In Section \ref{sec:multiwavelength} we describe existing multi-wavelength data in Chandra X-ray, HST optical, Spitzer infrared, and multi-instrument, multi-band radio interferometry of the BCG that are used to support our work.
We present the results from our SITELLE data in Section \ref{sec:results}, and discuss their analysis and the possible interpretations along with the multi-wavelength data in Section \ref{sec:Discussion}.
Our data, methods, and findings are summarized in Section \ref{sec:conclusions}.
Throughout this work, we assume a $\Lambda$CDM cosmology, with $\Omega_m=0.3$, $\Omega_{\Lambda}=0.7$, and H$_0$=70 km/s/Mpc. At the redshift of A2390, one arc second is equivalent to 3.65 kpc.
\begin{table*}
	\centering
	\caption{SITELLE Observation Details of A2390}
	\label{table:observations}
	\begin{tabular}{ccccccccc} 
		\hline
		Proposal ID & Filter & Wavelength Range (nm) & Seeing (\arcsec) & N$_{steps}$ & $\Delta\lambda$ (\AA) & R & Emission Lines (\AA) & Exposure (s) \\
		\hline
		16BC03 & C1 & 385-490 & 1.3 & 304 & 5.8 & 600 & \OII3727 (blended) & 12,160 \\
19BT07 & C2 & 562-625 & 1.2 & 297 & 3.6 & 1300 & \hbeta, \OIII4959,5007 & 13,365 \\
17AD94 & C4 & 798-823 & 1.3 & 124 & 5.2 & 1080 & \Halpha, \NII6548,6583 & 11,408
	\end{tabular}
\end{table*}

\section{SITELLE Observations of A2390} \label{sec:sitelleobs}

\subsection{Observations and Data}
\label{sec:cfhtsitelle}

\begin{figure*}
     \centering
     \includegraphics[width=\textwidth]{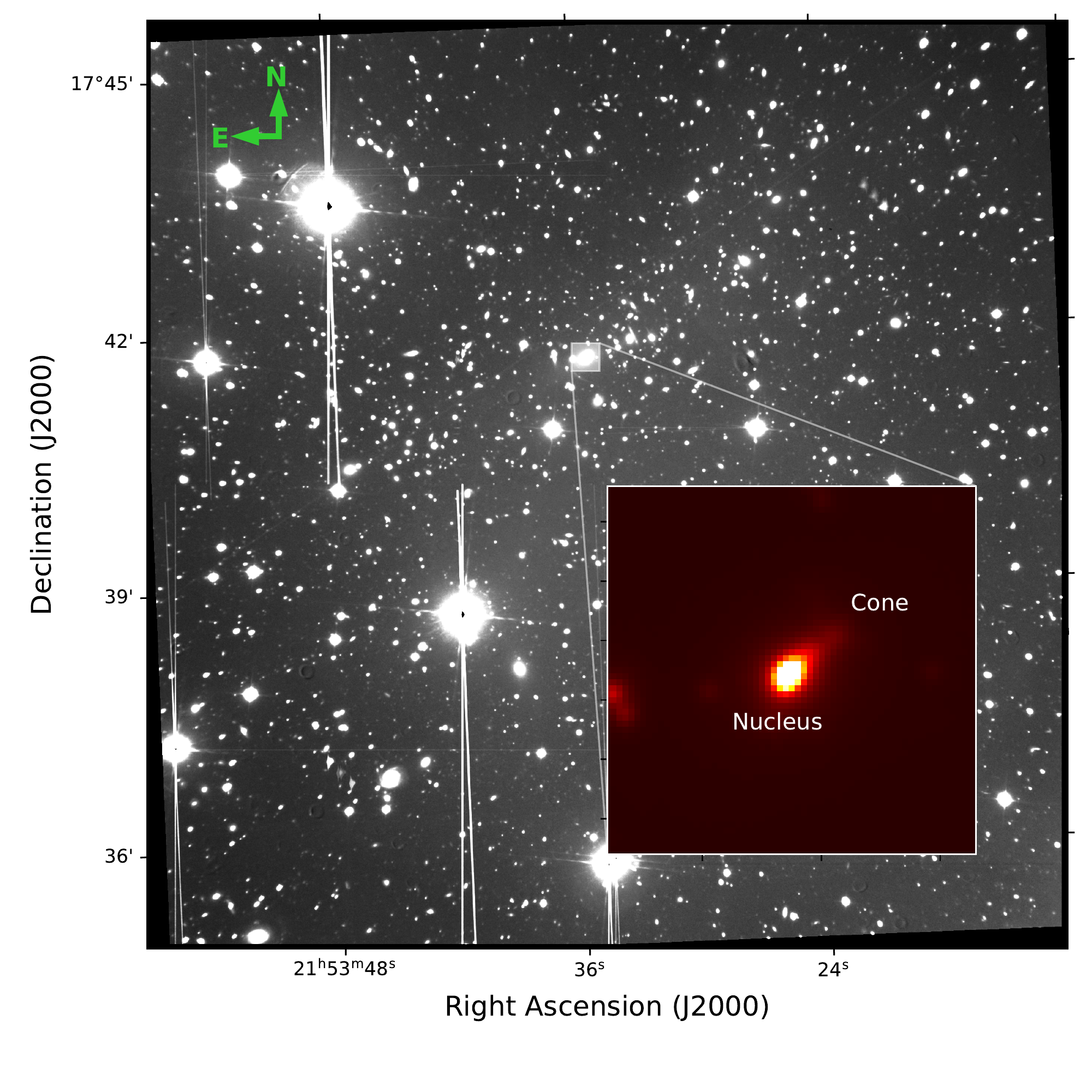}
        \caption{SITELLE deep image of the central pointing of A2390 ($z=0.228$). This image shows a single SITELLE pointing $11'\times 11'$ in size. The inset and zoomed image is the location of the brightest cluster galaxy (BCG), Object 1757C ($z=0.231$) from the catalog \citet{Liu2021}. The inset is a $10\arcsec \times 10\arcsec$ cutout under a different stretch to emphasize the nuclear region in contrast to the cone region of the BCG}. Several bright foreground stars show saturation streaks.
        \label{fig:deepimage}
\end{figure*}

SITELLE was originally intended for the study of emission line objects such as local nebulae, supernova remnants, and the HII regions of nearby galaxies as well as star-formation and emission line galaxies in galaxy clusters \citep{Drissen2019,RousseauNepton2019}.
The wide 11'$\times$11' field of view of SITELLE/CFHT makes it an ideal instrument for studying nearby galaxy clusters (see Figure \ref{fig:deepimage}).
By observing rich clusters such as A2390 with SITELLE, one can obtain imaging spectroscopy of an unbiased, luminosity-weighted sample of cluster galaxies including the cD BCG, identified as Obj1757C in the catalog generated in \citet{Liu2021}. The proper name for Obj1757C is J21536+1741 (from SIMBAD), but we use the catalog name for the BCG for brevity.

We obtained SITELLE imaging Fourier Transform Spectrograph (iFTS) data cubes in the C1, C2, and C4 band-limiting filters, which at the redshift of the cluster will detect the \OII\ doublet from $z=0.03-0.31$ (albeit blended), \OIII\ and \hbeta\ from $z=0.15-0.25$, and \Halpha\ and \NII\ from $z=0.21-0.25$, respectively.
These filters cover the primary optical diagnostic nebular emission lines at the redshift of Obj1757C.
Images from SITELLE have a spatial scale of $0.322\arcsec$/pix, which at the cluster redshift of $z\sim0.228$ is an angular scale of 3.6 kpc/$\arcsec$, or 1.17 kpc per pixel.
These data are available on the Canadian Astronomical Data Centre website at the observation IDs\footnote{https://www.cadc-ccda.hia-iha.nrc-cnrc.gc.ca/en/} listed in Table 1.
C1 was observed on August 27th, 2016 (PI: Loubser) as part of the SITELLE commissioning dataset.
C2 and C4 were observed on September 27th, 2019 (PI: Hsieh) and July 6th, 2017 (PI: Yee), respectively.
Exposure times (including overhead between wavenumber steps), resolution, seeing, and filter band widths are listed in Table 1.

\begin{figure*}
     \centering
     \includegraphics[width=\textwidth]{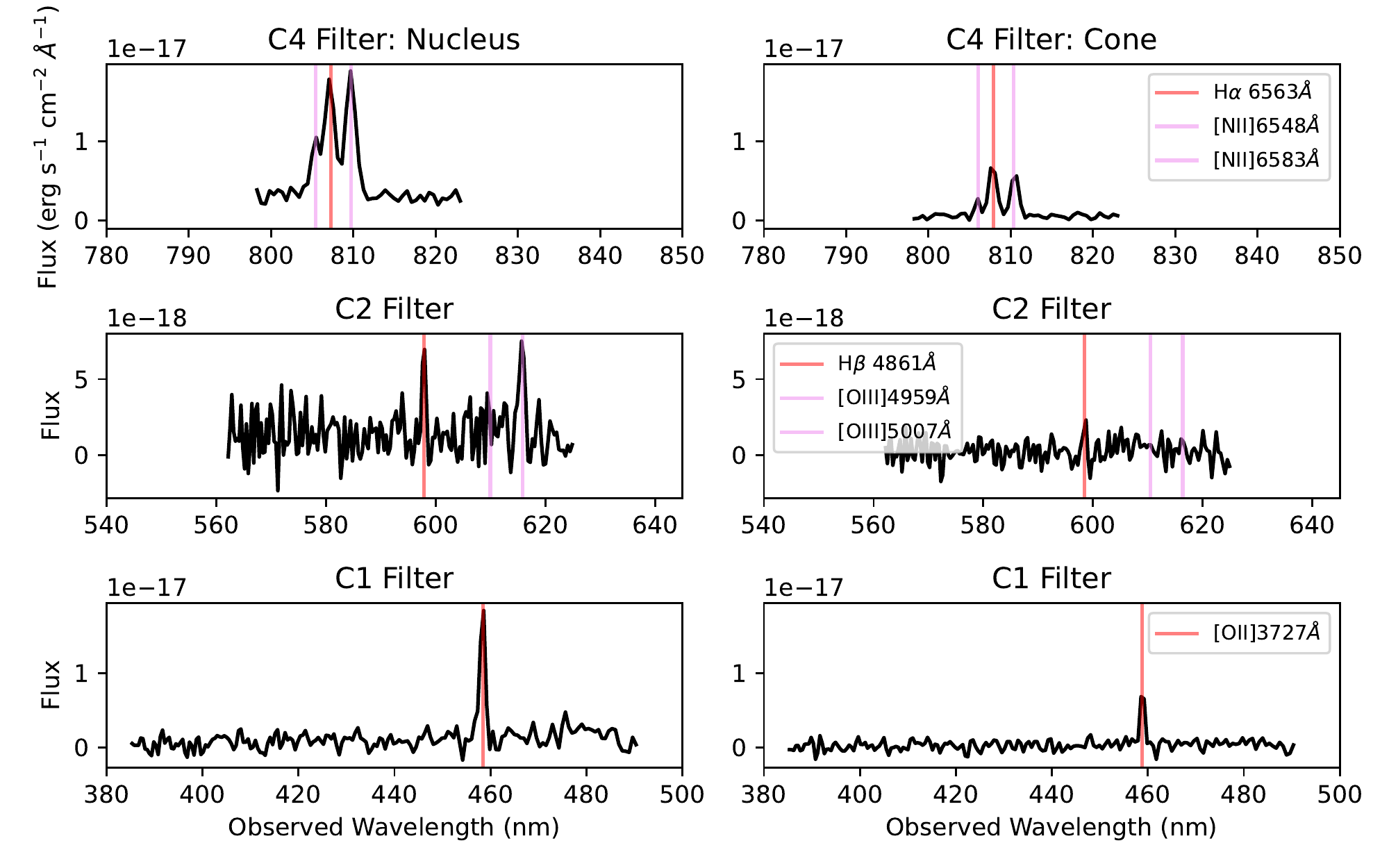}
        \caption{Example spectra at two single pixels in different locations of the BCG. The ``Nucleus" (left) column contains spectra in C4, C2, and C1 of the BCG at the location of the \halpha\ centroid. The ``Cone" (right) column contains spectra from the same pixel in the cone at 3.4$\arcsec$ from the \halpha\ centroid. In each panel, the spectrum is shown in black, and flux units are in \fluxunits\AA$^{-1}$. Emission line locations are marked in red and pink. We only plot spectral regions within the usable band-limiting filter channels.}
        \label{fig:spectra}
\end{figure*}

The initial data products are data cubes, with a spectrum at each spatial element (Figure \ref{fig:spectra}).
Each spectrum is made up of multiple ``channels", or wavelength steps.
The data are reduced and calibrated through the ORB software \citep{Martin2016}\footnote{https://github.com/thomasorb/orbs}.
The C4 data is processed through a software pipeline\footnote{https://github.com/NGC4676/SITELLE\_ELG\_finder} described in \citet{Liu2021}, which identifies \halpha-emitting galaxies and removes sky contamination.

\subsection{Emission-line Galaxy Identification and Cataloguing}
\label{sec:elgfinding}

Initial measurements of Obj1757C are generated from the cataloging algorithm described in \citet{Liu2021}. The software pipeline identifies emission line objects in SITELLE data cubes by fitting emission line templates for the \Halpha\ and \NII6548,6583\AA\ doublet, the \hbeta\ and \OIII4959,5007\AA\ doublet, and a single undefined emission line over the spectral coverage (channels).
The \citet{Liu2021} catalogs contain the centroid, ellipticity, radius, and position angle (PA) of the \Halpha\ emission-line region derived from the specific spectral channels that contain the \halpha\ line feature.

We summarize the processing steps of \citet{Liu2021} here, which resulted in the catalog used for this work.
The sky background for each channel is evaluated via a $128\times128$ pix$^2$ box using its mode, and this value is subtracted from the channel.
A low-pass filter is applied to the data cube to minimize fringing, which is produced by the combination of the interferometric nature of SITELLE and the presence of some bright sky lines.
A stacked image is created by summing along a restricted set of wavenumber channels which have good transmission, and the stack is used to detect sources with the photutils Python package.
Emission line galaxies are identified by subtracting smoothly-varied continua, and then matched to emission line templates using a cross-correlation technique (see Section 3.6 of \citealt{Liu2021} for details).
The cross-correlation measures the best-fit $z_{spec}$ and emission line identities of each measured galaxy.

\subsection{Spatially Resolved Emission Line Mapping}
\label{sec:lucifits}

\begin{figure*}
\centering
\includegraphics[width=\textwidth]{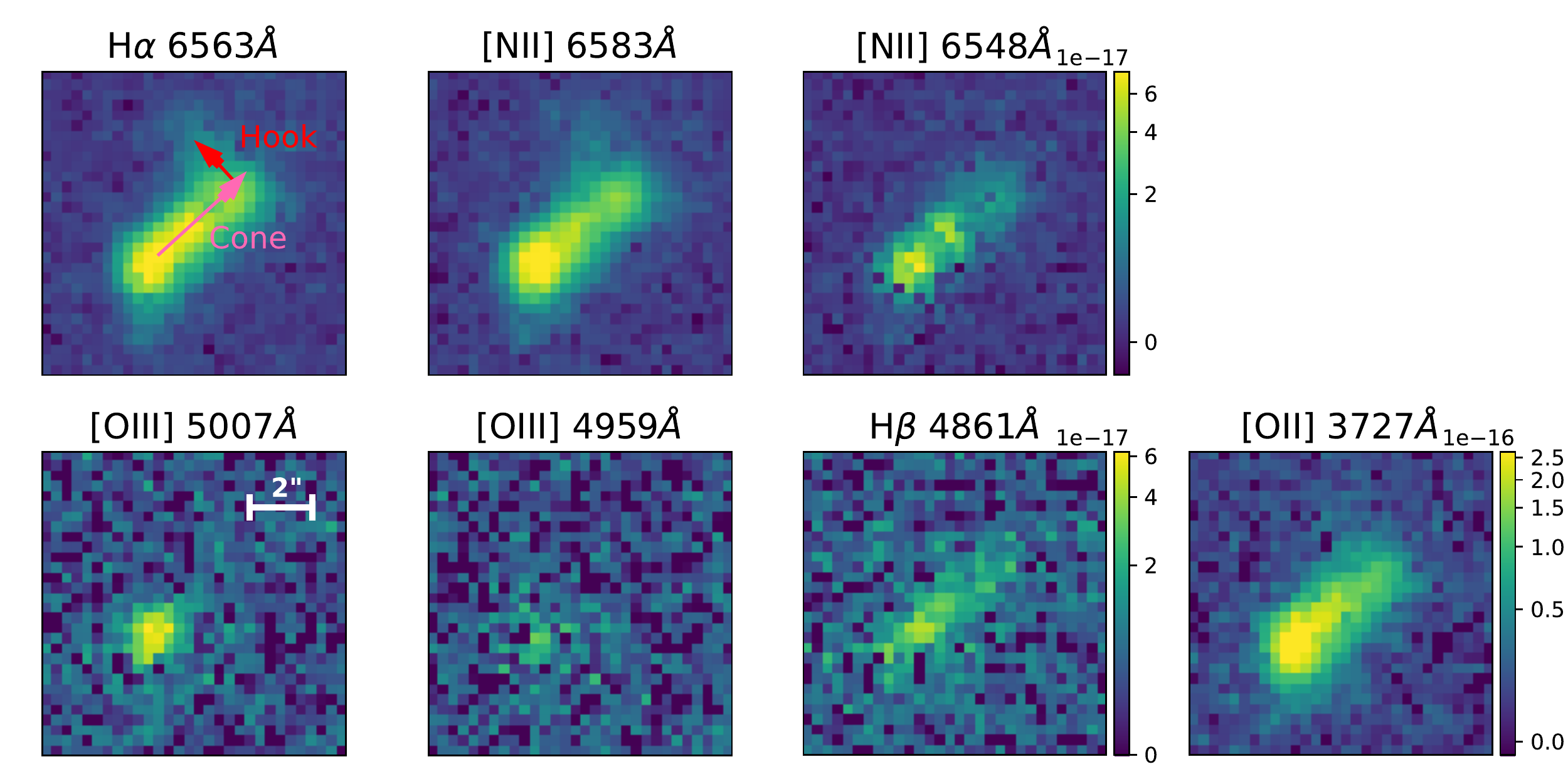}
\caption{Flux maps of emission lines derived from LUCI. The flux (given by the colorbar) is in units of \fluxunits\AA$^{-1}$. Only three colorbars are shown because flux maps from each filter use the same colormap (i.e. \halpha\ and \NII\ from C4, \OIII\ and \hbeta\ from C2, and \OII\ from C1). Top row from left: The flux maps derived from the C4 SITELLE filter, which provides coverage of the redshifted \halpha\ and \NII\ emission lines. These maps show weaker \NII\ emission in the cone and hook and bright emission in the core. Bottom row: The farthest left three panels are flux maps derived from the C2 filter, covering the redshifted \OIII\ and \hbeta\ emission lines. The farthest right panel is the \OII\ emission line doublet from the C1 filter. We see very strong \OII\ emission compared to the \OIII\ and \hbeta\ lines, indicating low ionization.}
\label{fig:fluxmaps}
\end{figure*}
\begin{figure*}
\centering
\includegraphics[width=\textwidth]{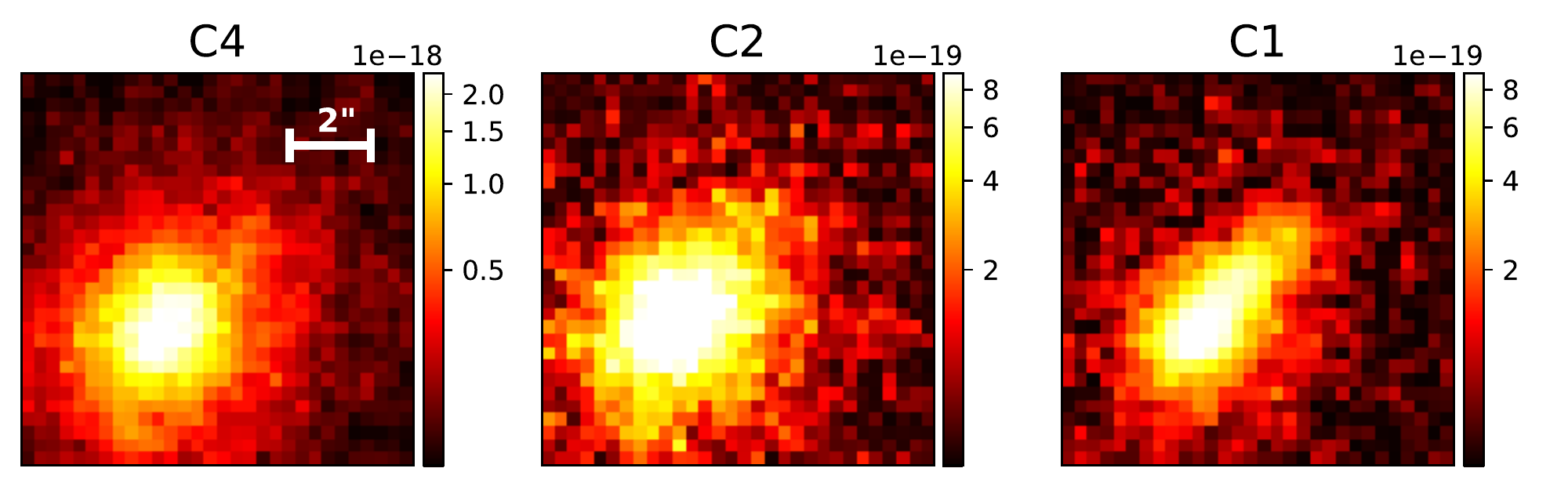}
\caption{Continuum maps of C4, C2, and C1 SITELLE filters derived from LUCI, with all emission line flux removed. The flux is in units of \fluxunits\AA$^{-1}$, same as in Figure \ref{fig:fluxmaps}. The C4 filter (rest-frame 6482\AA\ to 6686\AA) displays only a mild elliptical shape of $\epsilon=0.29$ and no cone. The continuum of the C2 filter (rest-frame 4565\AA\ to 5077\AA) is more elliptical than C4 and has no cone. The C1 filter which covers rest-frame 3128\AA\ to 3980\AA\ is effectively the rest U-band continuum, and is elongated in a direction consistent with the emission line cone. }
\label{fig:continuummaps}
\end{figure*}

Using LUCI\footnote{https://crhea93.github.io/LUCI/index.html} \citep{Rhea2021a,RheaLUCI2021}, a set of programs specifically designed to analyze iFTS spectral imaging data such as those from SITELLE, we can extract maps of the redshifted emission line flux.
These steps are performed for each observed filter of Obj1757C (C1, C2, and C4).
From the \citet{Liu2021} catalog, we input the redshift and spatial position of the estimated \Halpha\ emission profile centroid, and fit emission line profiles and continuum values over individual pixels covering a square region 4$\times$ the galaxy's \halpha\ equivalent radius\footnote{From the Python package photutils \citep{larry_bradley_2020_4044744}, the radius of a circle with an area equal to the source segment on the segmentation map.} on each side.
Pixels in this region are fit independently by LUCI.
We subtract background sky levels from an empty circular aperture near the galaxy 5.9\arcsec\ in radius, which in the case of Obj1757C was located $18\arcsec$ away southwest from the \halpha\ centroid.

LUCI fits the velocity-shifted position of the emission line, which provides a velocity measurement relative to the input object redshift derived from an integrated spectral fit.
LUCI measures the width, or broadening, of the emission line via a SincGauss function, which is the convolution of a sinc function with a Gaussian function (for broadening the sinc function) and describes the instrumental line shape for extended objects as observed by SITELLE \citep{Martin2016}.
The extracted flux maps are fit over all of the spectral channels to find the location of an emission line, and therefore any detected emission will be included in the map despite a velocity shift or kinematic broadening.
Additionally, LUCI measures the continuum of the galaxy spectrum simultaneously with the line flux in the measured pixel, therefore providing a continuum-subtracted emission line flux.
Continuum values are obtained by fitting a flat continuum with zero slope, simultaneously with the SincGuass function of the emission line, to the spectrum at each pixel.
The filters C1, C2, and C4 are all narrow band-limiting filters of under 110 nm; visual inspection confirms a flat continuum over most pixels after background subtraction.

Initial guesses for line broadening due to velocity dispersion and line shift due to velocity are provided from a training set created in a similar method to that described in \citet{Rhea2021}, which we briefly summarize here.
Reference spectra for C1, C2, and C4 are created for each filter over the wavelength coverage of each filter.
We redshift the emission lines to the redshift value of Obj1757C ($z=0.231$, determined from the integrated spectrum of the galaxy described in \citet{Liu2021} and Section \ref{sec:elgfinding}).
We use line ratios for \OII3726,3729\AA, \hbeta, \OIII4959,5007\AA, \NII6548,6583\AA, and \Halpha\ taken from the Mexican Million Model database replicating HII regions \citep{Morisset2015}.
A set of 50,000 test spectra are generated with velocity shifts $v=$ -1000\kms\ to 1000\kms, ionized gas velocity dispersion \sigmagas$=0-300$\kms, and signal-to-noise ratio SNR $=2-30$.
An artificial neural network is trained on these data, separating them into a training and test set to determine the ability of the network to recover input values.
When a spectrum is inputted to LUCI, the network predicts the initial guess for the velocity shift and \sigmagas\ broadening values of the emission lines.
The fit is run through an MCMC algorithm with priors clustered around these initial guesses of velocity, \sigmagas, continuum levels, and amplitude of the emission lines to determine errors on the fit at each pixel.

Emission lines observed in each filter are fit simultaneously and are tied at a fixed velocity for each emission line complex.
\Halpha\ and the \NII\ doublet emission lines are allowed to vary in the line broadening (as well as the \OIII\ doublet and \hbeta) but the fit does not produce significantly different results if each emission line is fixed to the same broadening value for all emission lines.
The \OII\ doublet is fit as a single line because it is not resolved at the SITELLE C1 filter resolution.
The \OIII\ doublet, \OIII4959\AA\ and \OIII5007\AA, are held to a constrained flux ratio of 1:2.98.
Additionally, \NII6548\AA\ and \NII6583\AA\ are constrained to a fixed flux ratio of 1:3 \citep{Dojcinovic2022}.

The results of the LUCI software on our SITELLE data include maps of the emission line fluxes seen in Figure \ref{fig:fluxmaps} and discussed in Section \ref{sec:structure} (later used to derive flux ratio maps described in Sections \ref{sec:ionization} and \ref{sec:sfshockionization}), continuum emission maps seen in Figure \ref{fig:continuummaps}, and velocity and broadening maps discussed in Section \ref{sec:kinematics}.
Fractional uncertainty levels in each of the flux maps and continuum maps is low in the nuclear region (0.05-0.1 fractional error) and increases outward in the cone and toward the hook (0.1-0.15 in \halpha\ and \NII, 0.15-0.35 in \OIII, 0.1-0.2 in \hbeta, and 0.5-0.2 in continuum).

\section{Multi-wavelength Data on A2390} \label{sec:multiwavelength}

\subsection{Chandra X-Ray Imaging}
\label{sec:chandra}

\begin{figure*}
     \centering
     \includegraphics[width=\textwidth]{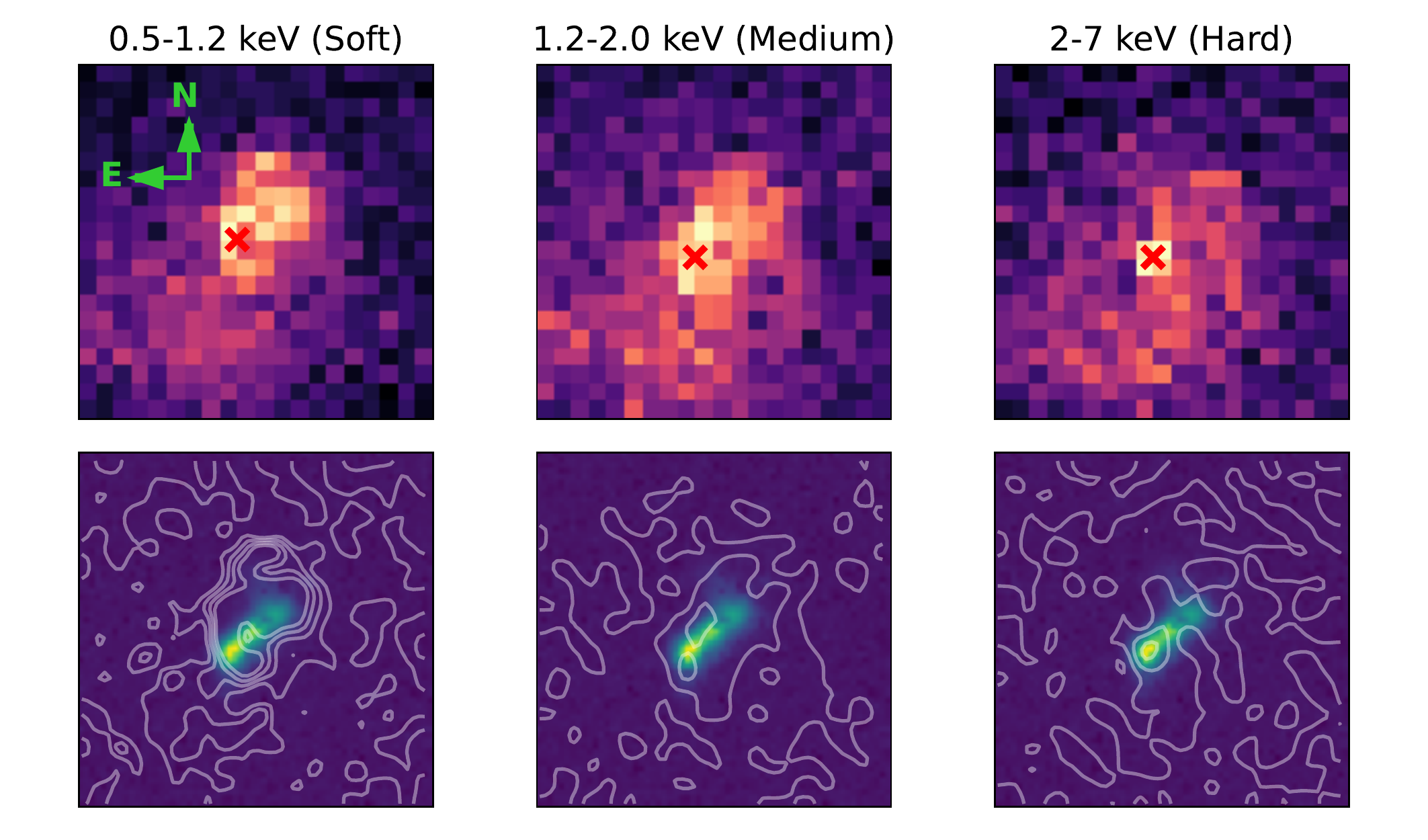}
        \caption{Chandra X-ray imaging of A2390. The top row are cutouts of the Chandra images in a $10\arcsec \times 10\arcsec$ square. The \halpha\ centroid location is marked with a red 'X'. The \halpha\ images with Chandra X-ray contours are shown in the bottom row. Excess in the  soft channels (left panels) appear to be located in the same space as the \Halpha\ cone, indicating cooler gas temperatures. The hard channels (right) are brightest at the galaxy center, the location of the AGN. }
        \label{fig:chandra}
\end{figure*}

Chandra X-ray imaging of A2390 was obtained by \citet{Allen2002} in hard (2-7 keV), medium (1.2-2 keV), and soft (0.5-1.2 keV) X-ray channels.
The Chandra observation (OBS ID: 4193) obtained from the archive consisted of a 96 ks ACIS pointing of the cluster center.
The exposure-corrected flux images were generated using the archive pipeline processed event file with CIAO v4.14 tools and CALDB v4.9.6 calibration files.
The pipeline processed Level 2 event file was used in conjunction with \emph{dmcopy} to extract the events and form images in different energy ranges.
The images were further binned 2x2 pixels to increase the signal-to-noise ratio.

The hard X-rays peak at the BCG nucleus (Figure \ref{fig:chandra}), and are associated with the AGN in the galaxy's center.
Diffuse soft emission is asymmetric and elongated along the extended emission line region (X-ray PA=46.1$^o$ from West) and the \halpha\ cone (\halpha\ PA=42.7$^o$ from West), similar to many \halpha\ filaments observed in BCGs which host buoyantly rising bubbles \citep{Fabian2003,McDonald2010, McNamara2016,Calzadilla2022}.
This elongated region is associated with a cooler temperature along the NW region of the ICM, in the same location as the \halpha\ emission line cone (see Figure \ref{fig:chandra}).

These data from the Chandra X-ray Observatory were also utilized by \citet{Sonkamble2015}, whose analysis identified five X-ray deficient cavities which appear around the BCG within 30\arcsec.
There is an edge to the 0.5-7 keV diffuse X-ray gas along the NW direction at 68\arcsec, or 246 kpc, associated with cold fronts \citep{Sonkamble2015}.
Intriguingly, the NW X-ray edge is possibly associated with the NW elongated region of \halpha\ flux and the elongated soft X-ray emission.
Buoyant gaseous bubbles are associated with the appearance of X-ray cavities like the ones seen in this analysis.

\subsection{HST Optical and Spitzer IR Data}
\label{sec:hstspitzer}

\begin{figure*}
\centering
\includegraphics[width=\textwidth]{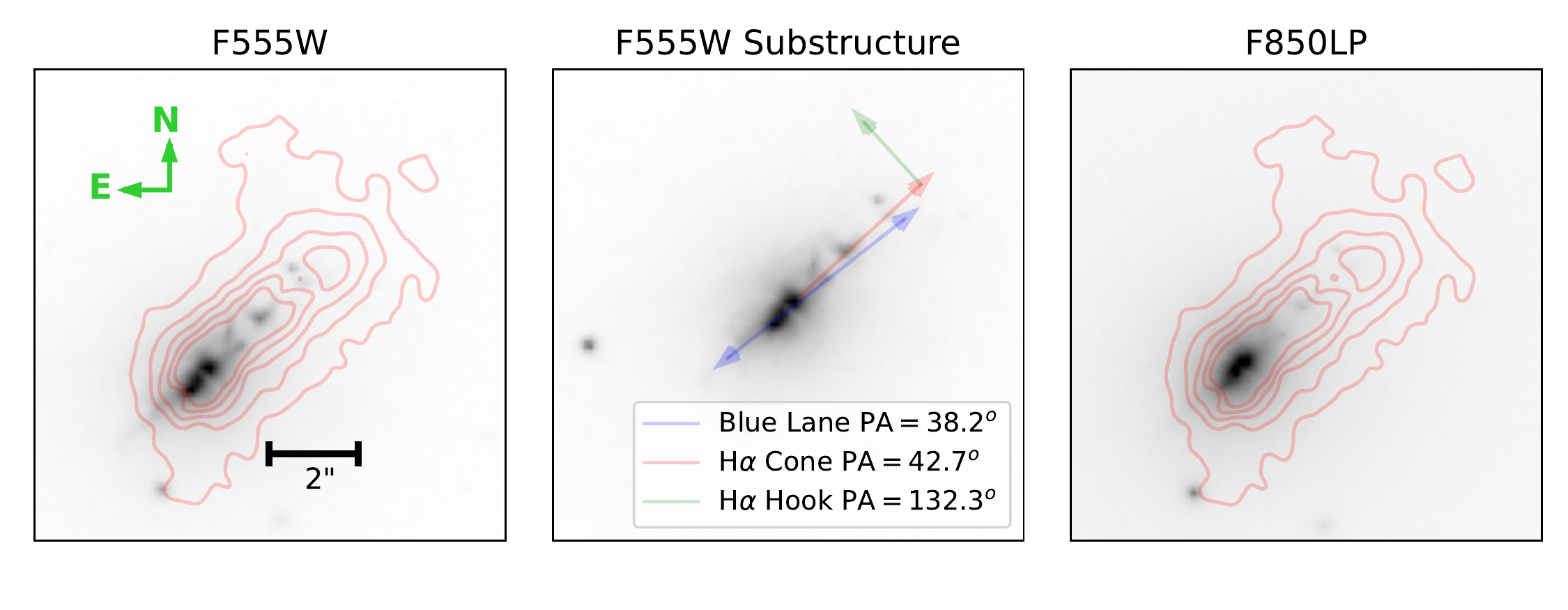}
\caption{Left: The HST/WFPC2/F555W image of the A2390 BCG. A two arcsecond bar is shown as a scale. Striking and complex substructure not resolved in the SITELLE flux maps is seen, including an extended region to the southeast and a gap, most likely a dust lane. \Halpha\ flux contours are overlaid in red. Center: The same image as left, but shifted upward and with structure directions overlaid. We include the narrow HST blue lane extending from the southeast side of the nucleus to the northwest (blue line), and the \halpha\ cone (red) and hook (green) as detected by SITELLE. The HST extended regions/blue lane emanating from the nucleus and the \halpha\ cone have position angles misaligned by $\sim4^o$. The most prominent clumps in F555W are along the \halpha\ cone. Right: The HST/ACS/F850LP image of the BCG, which also displays a less prominent dust lane and some of the substructure seen in F555W as one might expect due to less reddening. }
\label{fig:hst}
\end{figure*}

\begin{figure*}
     \centering
     \includegraphics[width=\textwidth]{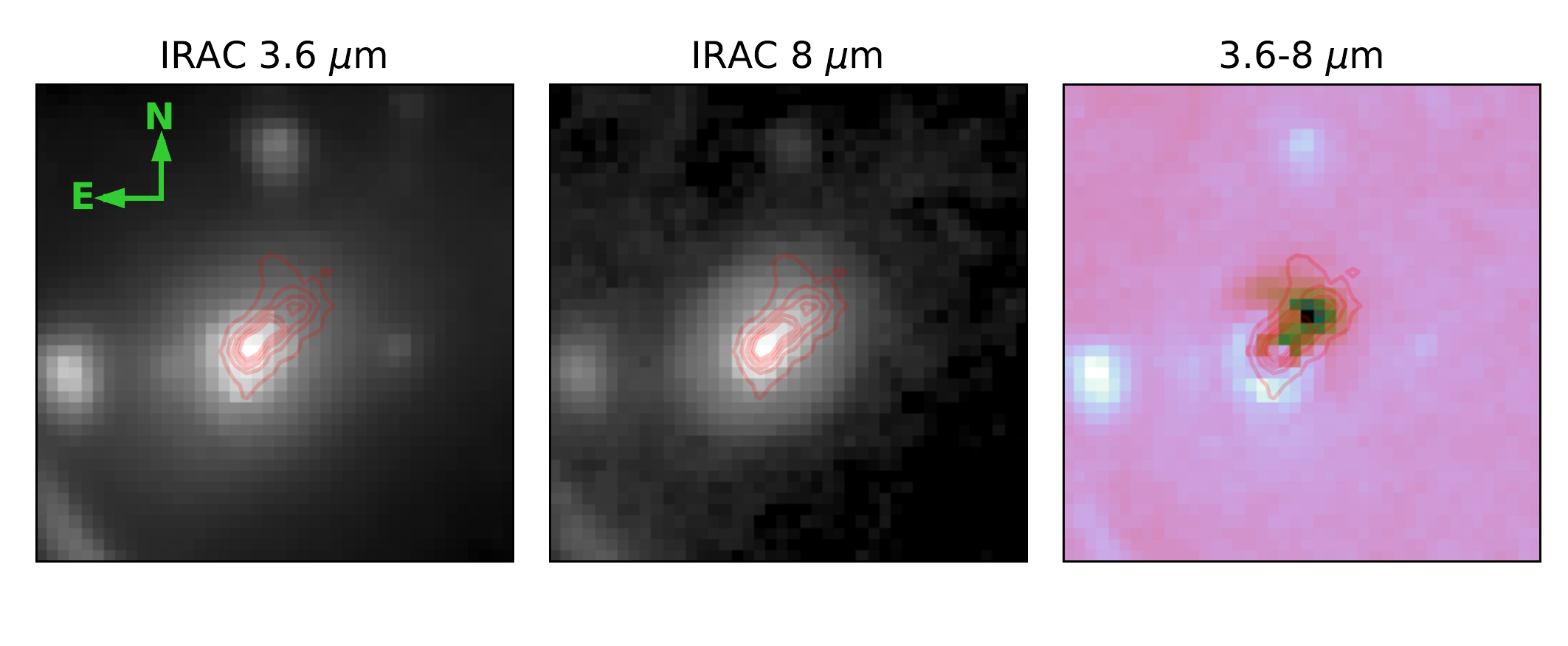}
        \caption{Spitzer/IRAC infrared imaging of A2390. Left: The 3.6\micron\ emission with \halpha\ flux map contours (Figure \ref{fig:fluxmaps}). The 3.6\micron\ channel is associated with the stellar population of the BCG and shows no extended structure. Center: The 8\micron\ emission, which has an extension towards the NW. Both images are normalized to the peak emission in each image. Right: Difference image of 3.6-8\micron. Regions dominated by 8\micron emission is represented by darker shades. The excess 8\micron emission region coincides with the \halpha\ cone which is shown by the red contours.}
        \label{fig:spitzer}
\end{figure*}

\citet{Hutchings2000} were the first to report a notably extended emission line region in the A2390 BCG.
They combine their CFHT/OSIS narrow-band \halpha\ imaging with HST/WFPC2 F555W, and HST/ACS F850LP images.
We obtained the same HST/WFPC2 F555W imaging \citep{Fort1994} for our own analysis from MAST, and performed no additional processing.
The extended HST continuum region was observed on both sides of the galaxy, but the southeastern side of the extended region is not observed in our SITELLE \halpha\ imaging or \citet{Hutchings2000} Ly$\alpha$ grism spectroscopy.
Knots of structure are seen close to the galaxy nucleus and along the emission line cone region in all observations.
Additionally, a possible dust lane within the central region of the galaxy is observable at a $90^\circ$ offset from the cone in both F555W and F850LP (see Figure \ref{fig:hst}).

Legacy Spitzer/IRAC IR imaging of A2390 in 3.6 (Channel 1) and 8 $\mu$m (Channel 4) were obtained using MAST \citep{Egami2006} (Figure \ref{fig:spitzer}).
The BCG of A2390 was the third IR-brightest BCG in the sample of \citet{Egami2006}, but it is not bright enough to be classified as a LIRG ($> 10^{11} L_{\odot}$).
A bright IR flux is correlated with a shorter radiative cooling timescale, which is common in cool-core clusters.
Based on IR spectral energy distribution signatures, the highly luminous IR emission was judged by the authors to be more likely caused by star formation rather than AGN emission.
8 \micron\ imaging would provide us polycyclic aromatic hydrocarbon (PAH) distribution in the BCG, and 3.6 \micron\ imaging provides IR imaging of the stellar component.
In the right panel of Figure \ref{fig:spitzer}, we created a difference image of the 3.6 and 8 \micron\ filters (left and middle panels, respectively) to emphasize the excess of the 8 \micron\ to the northwest.
We note that the 8 \micron\ emission is more extended than the 3.6 \micron, and the extended region is in the direction of the BCG cone.

\subsection{Radio Interferometry}
\label{sec:radio}

\citet{Augusto2006a} observed the BCG of A2390 (the radio source of the BCG is called B2151+174 in this study) using multi-frequency, multi-epoch radio interferometry from several arrays such as VLBA, the VLA, and MERLIN.
These data indicated that the BCG radio source has a complex structure, consisting of an extended mini-halo, and a number of compact mini-jets, with the two most prominent milli-arcsecond jets in the north-south orientation.
The authors also state that it has one of the flattest radio spectrums and most spatially compact radio sources known at the time of observation.
The 1.4 GHZ radio power of the source is $\sim10^{25.1}$ W Hz$^{-1}$.
The MERLIN 5.0 GHz map indicates a size of 0.49\arcsec\ ($\sim$1.5 kpc), making it a bright core medium-size symmetric object.

\citet{Savini2019} obtained LOFAR radio imaging of A2390.
They list the diffuse emission from low-resolution images as 1100 kpc in size.
The authors also identified two large extended lobes on the order of $\sim$600 kiloparsecs from the center region, separate from the giant radio halo.

\section{Results}
\label{sec:results}

\subsection{Emission Line and Continuum Structure}
\label{sec:structure}
\citet{Hutchings2000} noted the extended structure of the \halpha\ emission line in their CFHT/OSIS narrow band images, which the authors referred to as the ``emission cone"; we will continue this terminology.
We reconfirm the existence of an extended emission line cone in the BCG via the LUCI emission line maps (Figure \ref{fig:fluxmaps}).
The cone is observed in \Halpha, \NII6548,6583\AA, \hbeta, and \OII3727\AA\ at a position angle of 42.7$^o$ (from the West of the image) extending northwest of the nucleus of the galaxy.
The cone is not observed in \OIII4959,5007\AA.
The cone is observed in the C1 continuum but not in the C2 and C4 continuua (Figure \ref{fig:continuummaps}).
Additionally, the SITELLE maps can detect a North-Eastern protrusion from the cone, which we will refer to as the ``hook".
The hook is observed in the \halpha\ map as well as \NII6583\AA.
However, it is not observed in \OII3727\AA, \OIII4959,5007\AA, and \hbeta, in any continuum image including SITELLE, HST/F555W, and HST/F850LP imaging, or in the \halpha\ narrow-band imaging of \citet{Hutchings2000}.
Intriguingly, the hook is also correlated with soft X-ray structure observed with Chandra (Figure \ref{fig:chandra}, left panels).

At 1.8\arcsec\ from the galaxy nucleus, clumpy structure is seen in the cone in \halpha\ and \NII. These clumps are not visible in the other emission line maps.
The clumps are located in the same region as those found in the HST/F555W image (Figure \ref{fig:hst}).

The nuclear region displays a possible dust lane in the HST images.
The F555W imaging (Figure \ref{fig:hst}, left) has a gap visible between the center of the nucleus and the cone structure.
The F850LP imaging (Figure \ref{fig:hst}, right) contains the same overall structure as F555W.
The dimming between the two nuclear clumps is less pronounced in F850LP, as one might expect with lower reddening from dust.
Therefore the gap is likely a dust lane, rather than an indicator of a major merger close pair.
This possible dust lane is not observed in any SITELLE emission line flux map due to the resolution limits of ground-based non-AO imaging.

While no south-eastern extended region is seen in the \halpha\ flux maps, the resolution of HST provides a clear image of the extended region in WFC3/F555W and ACS/850LP (Figure \ref{fig:hst}), referred to in \citet{Hutchings2000} as the "blue lane".
The HST blue lane on the NW side and \halpha\ cone are offset by 4$^o$.
The significance of this offset is unclear due to the lower spatial resolution of ground-based SITELLE data compared to HST spatial resolution.

\subsection{Ionization and BPT Profiles}
\label{sec:ionization}

The  Baldwin-Phillips-Terlevich (BPT) diagram \citep{Baldwin1981} is a diagnostic figure that plots the relative flux ratios of emission lines to determine the state of the plasma or ionized gas.
The most widely used version of the BPT diagram is the \NIIHalpha\ vs. \OIIIHbeta\ ratio map, which can distinguish between photoionized vs shock ionized gas, as well as between LINERs and Seyfert type AGN.
This diagnostic method is very helpful for a galaxy such as Obj1757C, which has an unknown ionization source in the cone, and a known ionization source (a radio AGN) at the nucleus.
\begin{figure}
\centering
\includegraphics[width=0.45\textwidth]{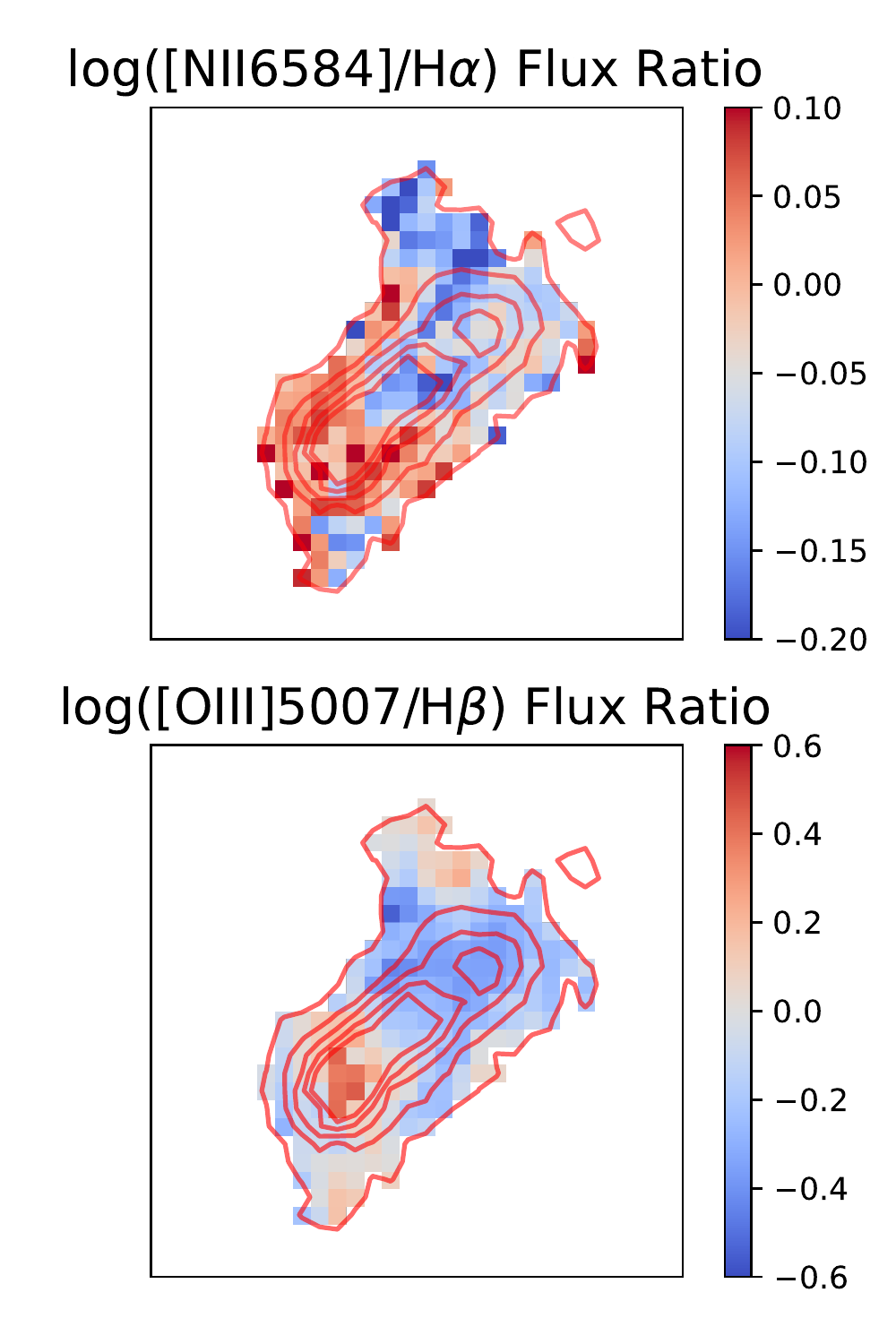}
\caption{Top: The log ratio of \NII6584\AA\ and \Halpha, a component of the BPT ionization diagnostic. This flux ratio map was smoothed using a 3X3 square filter to reduce noise in the low flux regions of the flux maps. The ratio appears relatively flat, but with a distinct drop between the nuclear region and the cone. This is consistent with other BCG measurements from \citet{Hamer2016}. Bottom: Log ratio of \hbeta\ and \OIII5007\AA. This ratio is also a component of the BPT diagnostic. This flux ratio map was smoothed using a 5X5 square filter to reduce noise in the cone, due to low flux values in the cone. The \OIII\ line is faint compared to other optical diagnostic emission lines, so it was not measurable over most of the galaxy. In both maps, the red contours are those of the \Halpha\ as seen in Figure \ref{fig:fluxmaps}.}
\label{fig:fluxratiomaps}
\end{figure}

\begin{figure}
\centering
\includegraphics[width=0.4\textwidth]{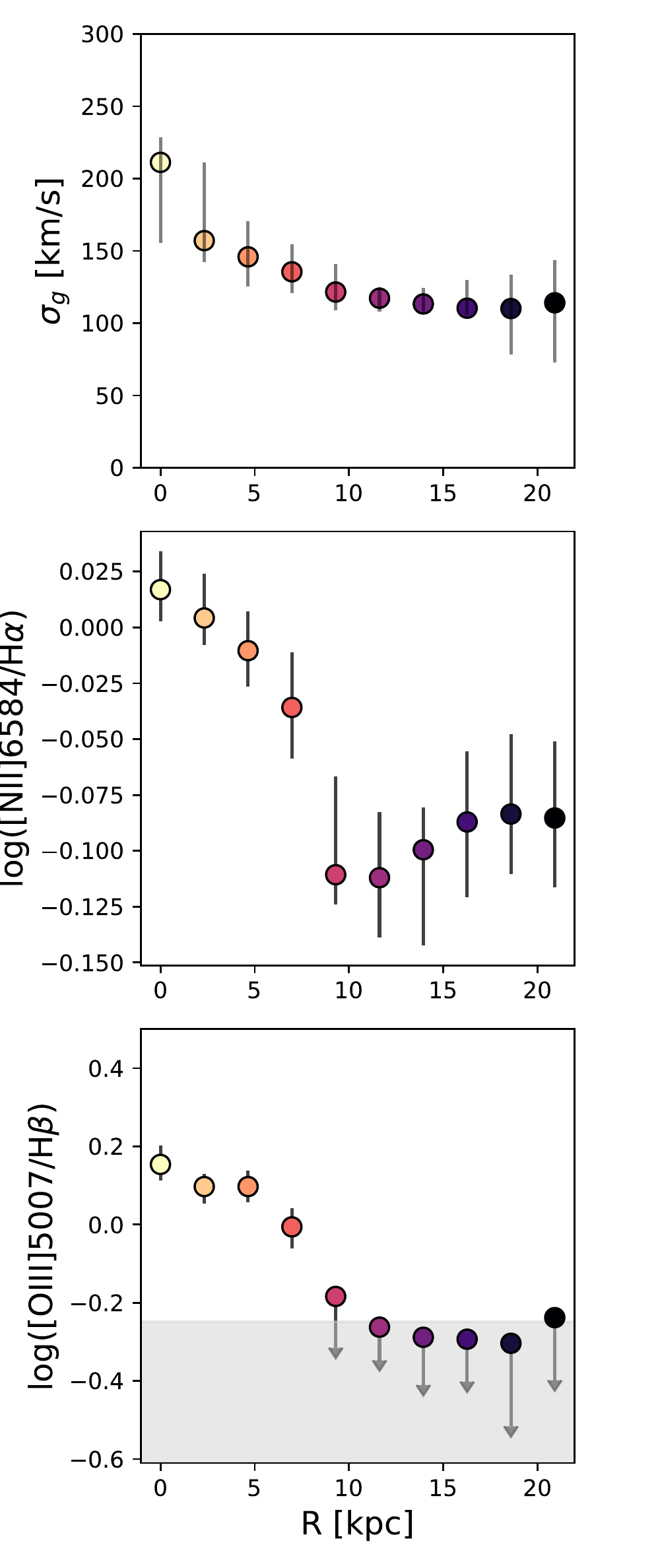}
\caption{Top: Median $\sigma_g$ derived from the \Halpha\ emission line fitting as a function of distance from the galaxy nucleus. The points are colored as a function of distance R from galaxy center (yellow) to the edge of the cone and hook (dark violet). These points are similarly colored for the middle and bottom panels. Errorbars in grey are the measurement errors derived from bootstrapping 1000 times per aperture. Each aperture is 1.0\arcsec\ in width along the cone and 2.3\arcsec\ in length. Middle: The median ratio of \NII6584\AA\ and \Halpha, a component of the BPT ionization diagnostic. Bottom: Median ratio of \hbeta\ and \OIII5007\AA\ as a function of distance from the galaxy center. This ratio is also a component of the BPT diagnostic. The shaded region is the detection limit of the flux ratio of a 2$\sigma$ detection of both emission lines. The \OIII\ emission line is not detected in the cone, making the points in the cone upper limits.}
\label{fig:fluxratios}
\end{figure}

To create ionization and BPT diagram maps, we use the LUCI emission line flux maps and flux error maps to create flux ratio maps (Figure \ref{fig:fluxratiomaps}).
We then extract overlapping rectangular apertures from the flux ratio maps along the direction of the cone from the galaxy center (as determined from the light-weighted \halpha\ ionized gas profile) to the edge of the cone (Figure \ref{fig:fluxratios}, lower two panels).
Each aperture is 1.0\arcsec\ in width along the cone and 2.3\arcsec\ in length, and overlaps with the previous and following apertures along the cone to smooth the resulting profiles.
We measure the median flux from the BCG nucleus to the cone.
Errorbars are derived bootstrapping the flux maps 1000 times.

Given that \OIII$5007\r{A}$ is not observed in the cone and is below the flux detection limit of $\sim4.26\times10^{-17}$ \surfacebrightnesslimit, the \OIIIHbeta\ flux ratio is an upper limit.
In contrast, the \NIIHalpha\ flux ratio is a secure detection until the edge of the cone.
Figures \ref{fig:fluxratios} and \ref{fig:bpt} show these flux ratio detections, using the same color scheme along the cone from lightest at the AGN core to darker along the cone.
Additionally, the 2$\sigma$ detectable limits on the flux ratios are displayed for the \OIIIHbeta\ ratio, where shaded gray regions are below SITELLE's flux detection limit.
The 2$\sigma$ detectable limits for \NIIHalpha\ are below the lower limit of the figure.
These flux ratio limits are calculated by measuring the surface brightness of the \halpha\ and \hbeta\ emission line structures in the cone, and determining the minimum flux ratio detectable at the SITELLE 2$\sigma$ flux detection limit.

\subsection{Kinematics of the \Halpha\ Ionized Gas}
\label{sec:kinematics}
\begin{figure}
\centering
\includegraphics[width=0.45\textwidth]{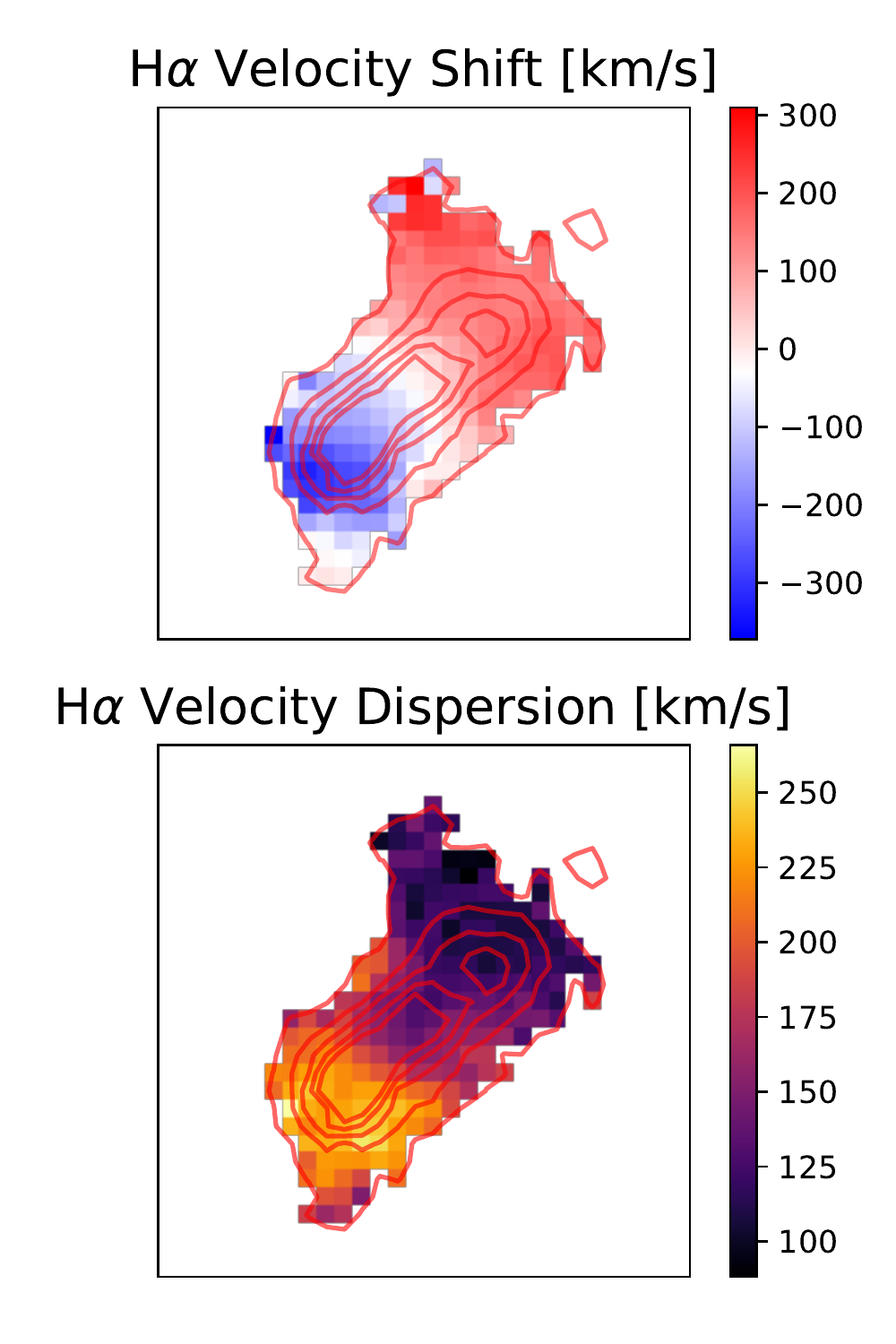}
\caption{Top: Velocity shift of the \Halpha\ line derived by LUCI. The velocity offset is with respect to the redshift of the integrated emission line gas from the \halpha\ and \NII\ emission lines ($z=0.231$) as observed in the integrated spectrum. Bottom: Gas velocity dispersion \sigmagas\ derived from the \Halpha\ line broadening. In both maps, the contours are the \Halpha\ flux contours seen in Figure \ref{fig:fluxmaps}.}
\label{fig:kinematics}
\end{figure}

We derive values for \Halpha\ velocity and \sigmagas, the velocity dispersion of the ionized gas, from LUCI by measuring the shift of emission lines from the integrated redshift of the emission line gas in Obj1757C ($z=0.231$), and from the broadening of the \Halpha\ emission line (Figure \ref{fig:kinematics}).
The velocity of the emission line gas has a 700\kms\ gradient from the nucleus to the cone and hook (Figure \ref{fig:kinematics}, upper).
The AGN region appears to be blueshifted and the cone and hook are redshifted relative to the luminosity-weighted integrated redshift of the galaxy's emission line gas.
No smooth velocity fitting is used (e.g. fitting to an arctangent rotation curve), but the velocity field is, nevertheless, remarkably ordered and smooth, indicating no recent disturbances due to major mergers, or a major merger with axes of rotation aligned.

The velocity dispersion profile peaks at the nucleus of the galaxy at 233 \kms, and decreases to $\sim130$ \kms\ in the cone and hook (Figure \ref{fig:kinematics}, lower).
The higher velocity dispersion region is associated with an AGN, but the lower \sigmagas\ region could be associated with rising buoyant bubbles of plasma blown by the AGN \citep{Aharonia2018,Zhang2022}.
These bubbles have low \sigmagas\ values because the motion of the gas is primarily dominated by outward jet stream motion (see Section \ref{sec:bubbles} for further discussion of plasma bubbles).
The clumpy substructure of the \halpha\ flux map and the HST/F555W imaging (Figure \ref{fig:hst}) is also associated with a slight flattening of the \sigmagas\ values (Figure \ref{fig:kinematics}, top panel), before an increasing decline in \sigmagas\ to the edge of the cone and hook.
Alternatively, the velocity dispersion values of the cone could be due to turbulence in the star forming regions, or due to slow shocks, or a combination of these reasons.

\section{Discussion} \label{sec:Discussion}

\subsection{Star Formation or Shock Ionization in the Cone}
\label{sec:sfshockionization}
\begin{figure}
     \centering
    \begin{subfigure}[b]{0.45\textwidth}
         \centering
        \includegraphics[width=\textwidth]{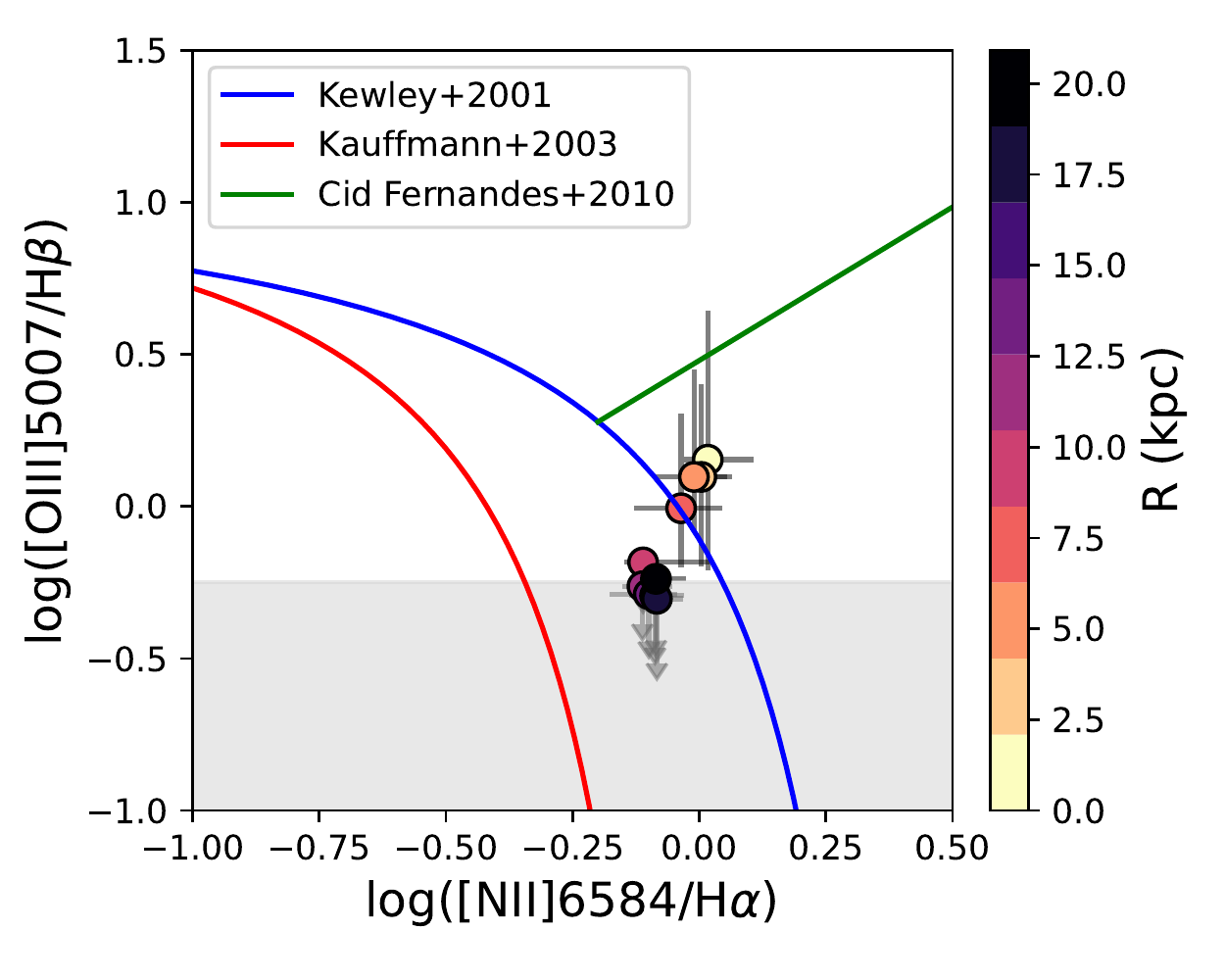}
    \end{subfigure}
        \caption{BPT diagram with points color coded as in Figure \ref{fig:fluxratios}, indicating distance from the galaxy center (yellow) to the edge of the cone and hook (dark violet). Errorbars are 2$\sigma$ range of values in each aperture as in Figure \ref{fig:fluxratios}. The red solid line marks the edge of the star-formation photoionization region, derived in \citet{Kauffmann2003}. The blue solid line is the edge of the composite ionization region from \citet{Kewley2001}. The green solid line is the dividing line between LINER and Seyfert galaxies empirically derived by \citet{CidFernandes2010}. The nuclear regions are located firmly in the LINER region, indicating shock-dominated photoionization with a soft ionizing field. However, the furthest points in the cone and hook region (dark violet points) move toward the composite photoionization region, indicating both shock and photo-ionization.}
        \label{fig:bpt}
\end{figure}

\begin{figure}
     \centering
     \begin{subfigure}[b]{0.45\textwidth}
         \centering
         \includegraphics[width=\textwidth]{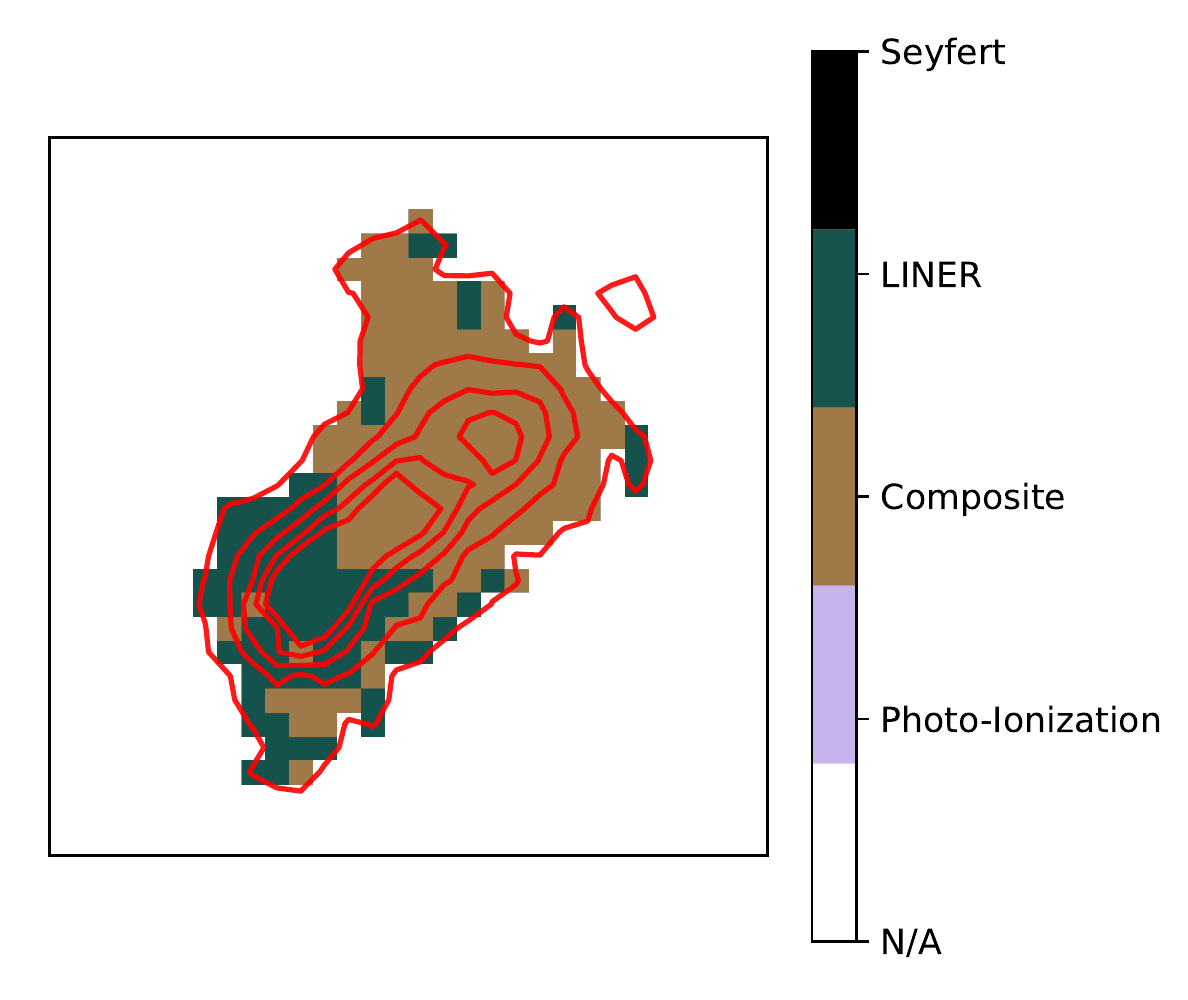}
     \end{subfigure}
        \caption{BPT map with pixels color coded by location on the BPT diagram (Figure \ref{fig:bpt}). The \Halpha\ and \NII\ flux ratio map was smoothed using a 3X3 square filter, and the \OIII\ and \hbeta\ flux ratio map was smoothed using a 5X5 square filter as shown in Figure \ref{fig:fluxratiomaps} to reduce noise. Regions without colored pixels were either outside the \halpha\ contours or contained an unusable flux for flux ratio analysis. The galaxy nucleus is dominated by LINER-like emission, powered by its AGN. The cone region is dominated by composite emission.}
        \label{fig:bptmap}
\end{figure}

The BPT diagnostics seen in Figures \ref{fig:bpt} and \ref{fig:bptmap} indicate that the cone is in a composite region; a combination of ionization state from photo-ionization and shock heating.
While emission lines can be the result of processes other than stellar photoionization \citep[e.g.][]{Li2020}, the clumpy structure of the cone and hook in \halpha\ emission line flux and HST imaging is possibly of a star-forming origin.
The galaxy nucleus houses an AGN, which is located firmly in the LINER region of the BPT diagram (Figure \ref{fig:bptmap}).
In Spitzer/IRAC imaging of the BCG, the 3.6\micron\ emission morphology is concentrated in the central AGN region (Figure \ref{fig:spitzer}, left).
In contrast, the 8\micron\ morphology is more extended to the northwest, into the same region as the \halpha\ cone (Figure \ref{fig:spitzer}, center), and could be associated with PAH emission from the star forming region \citep{Sivanandam2014}.
The soft X-ray emission (Section \ref{sec:chandra}) is also extended in the same orientation as the \halpha\ cone and the 8\micron\ emission (Figure \ref{fig:chandra}, left), and could be similarly associated with star-formation.

Similar star-formation activity in BCGs with AGN activity has been observed in the local universe up to $z\sim0.5$ \citep[e.g.][]{Crawford1999, McNamara2006, VonDerLinden2007, Fogarty2015, Hamer2016, Ciocan2021, maier_star-formation_2022}.
Our observations are consistent with BPT diagnostic measurements of BCGs in \citet{Hamer2016}, from VLT/VMOS IFU data of 73 galaxy clusters.
Note that the A2390 BCG was observed in that study, but its \halpha\ morphology was classified as a simple elliptical with centrally concentrated core, which is inconsistent with our own conclusion.
Instead, our observations of Obj1757C are more consistent with BCGs classified as ``Plumes" in \citet{Hamer2016}, where the \halpha\ emission shows a clear extent in one preferential direction not shared by the continuum.
These objects also showed LINER-like emission in the nuclear regions, and composite emission in the plumes, consistent with our own measurements of Obj1757C (Figure \ref{fig:bptmap}).

\Halpha, while commonly used as an indicator of star-formation \citep{Kennicutt1998}, can also be produced from other processes such as shock heating.
Based on the high \halpha\ emission, the clumpy structure observed in both HST imaging and the SITELLE \halpha\ maps, and the possible presence of PAH from IRAC 8um imaging, star formation could be occurring in the cone, but further data are needed to confirm the presence of star-formation.
While the source of ionization in the cone of Obj1757C is firmly in the composite region of the BPT diagram, it is not shock dominated, because \halpha\ \sigmagas\ values are low ($\sim100-200$\kms, see Section \ref{sec:kinematics}), which is inconsistent with \sigmagas\ values for fast shocks \citep{Fabian2012,Hamer2016,Aharonia2018}.
Slow shocks could be powering the ionization, but other processes such as the interplay between hot plasma and cold gas could fuel the ionization seen in the cone.
The soft X-ray plume seen in Figure \ref{fig:chandra} could be an indicator of this process \citep[i.e. ``turbulent diffusive reconnection"][]{Fabian2011}.

To confirm if star-formation is occurring in the cone, high-resolution observations of cold gas (CO features or H$_2$) such as from ALMA would be enlightening \citep{Russell2017}.
Obj1757C has been observed with ALMA, but the authors do not have access to these data currently.
Additionally, high-resolution \halpha\ imaging or IFU data would be useful, to determine whether the \halpha\ flux is coincident with the HST clumpy structure seen in Figure \ref{fig:hst}.

\subsection{Kinetic Mode AGN and ICM Bubbles}
\label{sec:bubbles}

Obj1757C displays characteristics quite similar to the ICM ``bubbles" \citep[e.g.][]{Fabian2003,Salome2006,McDonald2010,Pope2010,Fabian2012, Tremblay2015,McNamara2016, Chen2019b}.
\citet{Fabian2003} note that filamentary structures in the central BCGs of galaxy clusters are associated with \halpha\ emission and soft X-ray emission, both of which are consistent with what is seen in Obj1757C.

Additionally, A2390 is a cool-core cluster, and these types of clusters quite commonly show bubble activity \citep[e.g.][]{Fabian2012}.
Cool-core clusters have a feeding mechanism for powering an AGN jet mode.
Pristine cold gas from the cooling flow falls onto the BCG and accretes onto the AGN.
As the AGN is fueled by this cooling flow, its jets inflate bubbles of plasma on either side of the galaxy nucleus.
The mechanism for how bubbles rise and maintain their structure is not fully understood.
Once the bubbles have formed, one possible scenario includes that they separate from the jet and rise buoyantly through the hot ICM \citep{Churazov2000, McNamara2000}.
Models where the bubble is allowed to rise as a rigid body \citep[e.g. due to a magnetic draping layer as proposed in][]{Dursi2008} show that detachment of the bubble from the jet is due to eddies of thin gaseous structures, similar to the \halpha\ filamentary structures seen in observations of the Perseus cluster \citep{Fabian2003}.
The filamentary structures form from gas condensing along the path of a buoyantly rising bubble \citep{Pope2010, Gaspari2012, Sharma2012,Li2014b,Li2014, Voit2017}, or are lifted by the bubble itself \citep{Churazov2001, Revaz2008}.
The structures, after detaching, form a stream of gas falling back toward the cluster center as the bubble rises \citep{Gaspari2012,McNamara2016,Zhang2022}.
In cool-core clusters, the bubble is later disrupted by cold fronts \citep{Fabian2021}, which limits the number of bubbles observed in the outer ICM regions of cool-core clusters.

\subsection{Signatures of AGN Precession?}
\label{sec:precession}

\begin{figure*}
     \centering
     \includegraphics[width=\textwidth]{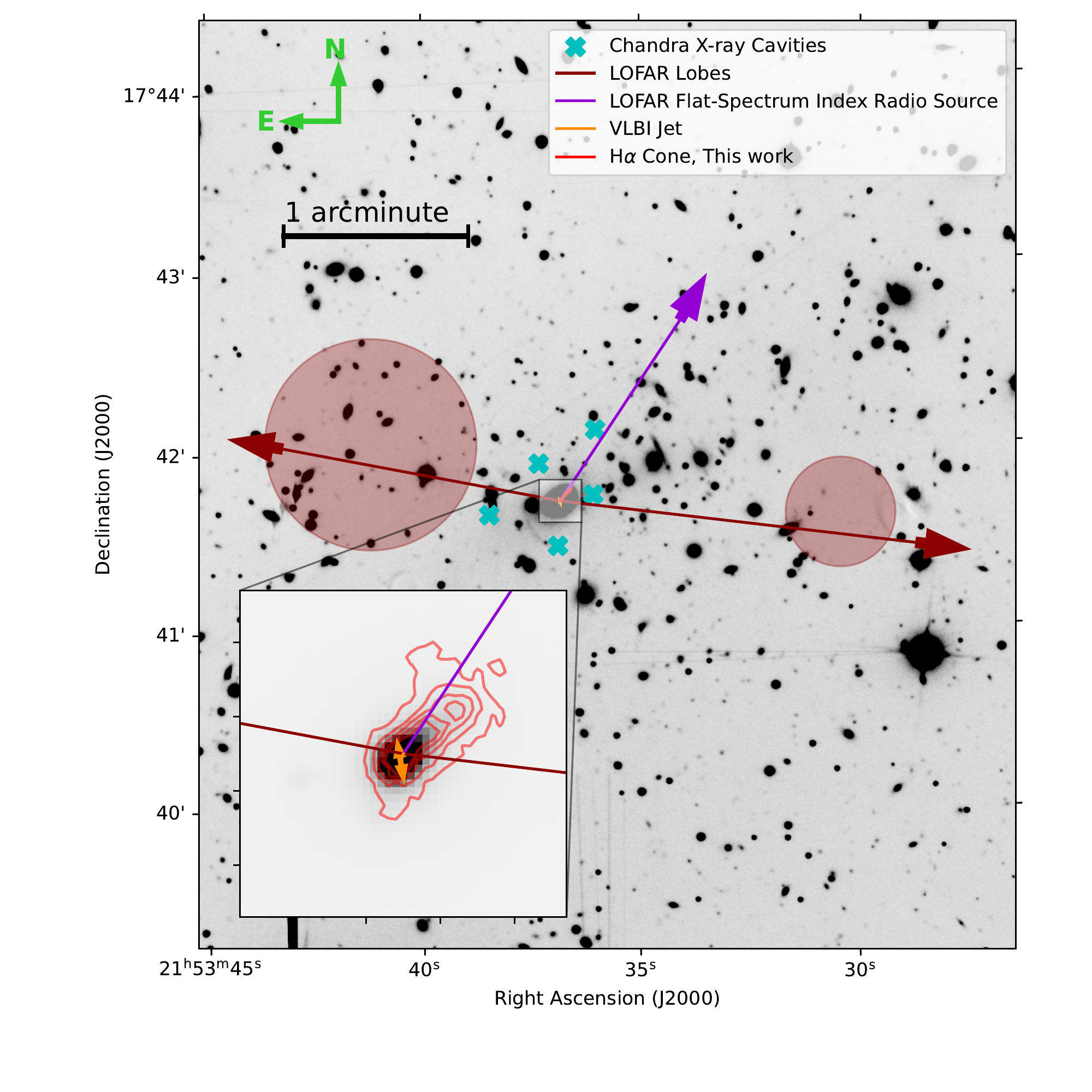}
        \caption{SITELLE deep image of the central pointing of A2390 ($z=0.228$) on which sources from radio,optical and x-ray observations are marked. Larger-scale radio lobes observed in \citet{Savini2019} by LOFAR are seen in dark red, extending out past $\sim$300 kpc. The sizes and locations of the red circles represent the sizes and locations of the radio lobe contours in \citet{Savini2019}. A region of flat spectral index radio emission not marked as a lobe is indicated by the violet arrow. These arrows mark the locations and fullest extent of the measured radio regions. Locations of five X-ray cavities measured in \citet{Sonkamble2015} are indicated with cyan crosses. The size of the crosses is not indicative of the size of the cavities. The inset is a $7\arcsec \times 7\arcsec$ cutout in which the milliarcsecond-scale north-south radio jets observed in \citet{Augusto2006a} are marked in orange. Note that the arrows are indicated for direction only and do not represent the size of the jet, which is smaller than is feasible to depict at this scale. The \halpha\ contours are shown in red. The radio lobes and jets radiate from the radio center of the BCG, and the \halpha\ jet radiates from the \halpha\ centroid measured as in Section \ref{sec:elgfinding}. Note that the position angle of the LOFAR flat spectral index radio emission region is close to those of the \halpha\ cone and one of the X-ray cavities.}
        \label{fig:precession}
\end{figure*}

Multiple episodes of AGN activity is suggested by multi-wavelength data in the literature and our SITELLE \halpha\ observations.
The evidence of multiple episodes is summarized and illustrated in Figure \ref{fig:precession}.
\citet{Augusto2006a} first suggested that there may be multiple episodes of precessed AGN activity in the A2390 BCG. They analyzed the (multi-wavelength, spatially resolved) radio morphology of the core region of the A2390 BCG using radio interferometry.
The authors find their MERLIN 5.0 GHz map indicates a compact size of 0.49\arcsec ($\sim$1.5 kpc) with a very young milliarcsecond-scale two-jet North-South structure in VLBI measurements.
We stress that the orange arrows in Figure  \ref{fig:precession} do not trace the extent of the north-south jets, but are in fact much larger than the milliarcsecond scale of the jets and are enlarged to emphasize the orientation with respect to other episodes of AGN activity.
\citet{Augusto2006a} point out  that  the jets  have  a $\sim45^o$ misalignment from the HST blue lane that extends from the South-east to North-west as seen by \citet{Hutchings2000} (see Figure \ref{fig:hst}).
The authors  suggest an interpretation of the misalignment as a result of a  previous episode of AGN activity in a different orientation.
Furthermore, the northern radio jet exhibits a twisted structure, similar to the hook structure of the \halpha\ cone  (Figure \ref{fig:fluxmaps}), which bends to the North-east,  suggesting there may be a precession of the radio jet.

\citet{Savini2019} analyze the large-scale radio structure of A2390 using LOFAR, noting an extended double lobe of $\sim600$ kpc.
We refer to these lobes as the east-west lobes, which are shown schematically in Figure \ref{fig:precession} as dark red arrows, with dark red circles indicating the location and extent of the widest section of the lobes.
The radio lobes of A2390 are larger than any other in their sample of 9 clusters, because usually the ICM prevents lobes this size from forming in clusters \citep{Wing2011,Savini2019}.
Structures with similar radio lobe scale include the BCG of cluster MS0735.6+7421 \citep[$\sim$600 kpc][]{mcnamara_energetic_2009,Vantyghem2014} and the BCG of the Ophiuchus cluster \citep[$\sim$570 kpc][]{Giacintucci2020}.
While \halpha\ is observed in these galaxies, in contrast to A2390 there is no extended nebular line structure associated with the BCG itself in either cluster \citep{mcnamara_energetic_2009,McDonald2010,Durret2015}.

\citet{Savini2019} argue that the lobes observed are remnants of an older AGN active phase, because the radio source at the BCG core contains a relatively flat spectrum, whereas the lobes have a steeper spectrum and appear to be fading.
They suggest a possible scenario consistent with that proposed by \citet{Augusto2006a}, is that there have been two epochs of AGN and outflow activities.
The older AGN activity is marked by the large, fading east-west lobes, and the newer AGN episode would be associated with the small-scale north-south jets.
However, we  note that \citet{Savini2019} did not consider the extended optical emission,  both continuum and emission line, found in the HST image with a PA in between those  of the radio jets and the radio lobes.

\citet{Savini2019} also includes a spectral index map of A2390 (their Figure 13).
There is a region of flat spectral index sources to the northwest of the BCG, which was not commented  upon by  the authors.
This region's direction and greatest extent is marked by a violet arrow in Figure \ref{fig:precession}.
This region is located at a similar orientation to the \halpha\ cone (see Figure \ref{fig:precession}), although is extended to $\sim1.5$ arcminutes ($\sim$275 kpc).
The flat radio spectrum region, along  with the \halpha\ cone, can  be interpreted as being associated with the same AGN activity.

The X-ray analysis of \citet{Sonkamble2015} confirms the presence of AGN feedback activity associated with the formation of buoyantly-rising bubbles from the soft X-ray spectra.
Additionally, the multiple X-ray cavities at varying orientations indicates that the direction of the feedback is changing with time.
There are four major X-ray cavities in the north, south, east, and west of the BCG, but \citet{Sonkamble2015} also observes a smaller cavity to the northwest of the BCG.
This cavity is located at a similar orientation to the \halpha\ cone observed by SITELLE and the flat spectral index radio region measured by LOFAR (see Figure \ref{fig:precession}).

Combining all these data, we propose a possible scenario of a third episode of AGN activity associated with the \halpha\ cone, with an age between those of  the episodes that  gave rise to  the large radio lobes and the milli-arcsecond  jets.
The emission line cone is possibly associated with an intermediate age episode of AGN emission that is lifting cooler ionized and molecular gas into the ICM, forming these X-ray deficient cavities.
An alternative explanation could be that the \halpha\ cone and the X-ray cavities and radio regions associated with it could be an example of one of these bubbles that has reached its maximum rising height, and the gas it entrained behind it has fallen back toward the center of the cluster \citep{Fabian2003}.
As it falls, it forms a nebular filamentary structure with star formation present.

All lines of evidence including radio and X-ray data point to Obj1757C harboring an extraordinary multi-episode active AGN, likely a kinetic-mode AGN, receiving a cold gas stream which is feeding the central SMBH.
The kinetic mode in cool-core clusters is associated with the formation of the bubbles seen in the Perseus cluster \citep{Fabian2003,Fabian2011}, but the ongoing question is whether this bubble formation happens in bursts or is a constant process that occasionally results in a detached bubble \citep{Fabian2012}.


\section{Conclusions} \label{sec:conclusions}

We present CFHT/SITELLE observations of the BCG (named Obj1757C) of the Abell 2390 galaxy cluster located at $z=0.228$.
Obj1757C displays a complex morphology, most notably an extended region of emission in HST/F555W, HST/F850LP, Spitzer/IRAC 8\micron, Chandra 0.5-1.2 keV and 1.2-2 keV, and now \halpha\ emission line imaging, referred to as the ``cone".
The origin of the cone was unclear due to multiple possible processes that could produce this emission line morphology.

To investigate the origin of the extended region, we create emission line flux maps of the blended \OII\ doublet (from the C1 filter), the \OIII\ doublet and \hbeta\ (C2), and the \NII\ doublet and \Halpha\ (C4).
The BCG emission line flux maps, continuum maps, and kinematic properties were extracted from LUCI software, which performs a Bayesian analysis of each pixel of a SITELLE IFU data cube, and extracts physical properties by fitting a SincGauss function to each emission line.

Emission line ratio maps are derived from our emission line flux maps, allowing us to perform a spatially resolved BPT analysis of the source of the ionizing radiation.
Emission line ratios show that the nucleus is dominated by LINER emission from the AGN, and that the cone is in a composite region of the BPT diagram that is fuelled by both photo-ionization and shock heating.
However, low values for \sigmagas\ show that the cone is not shock dominated, so fast shocks cannot be powering the ionization mechanism.
Slower shocks and the interplay between hot plasma and cold gas, and star formation regions could fuel the ionization as an alternative explanation.

Our results indicate that Obj1757C is consistent with being a kinetic-mode AGN that is forming \halpha\ filaments due to the inflation, detachment, and rising of buoyant plasma-filled bubbles.
The kinematic maps indicate that Obj1757C has a velocity gradient of 700 \kms\ between the nucleus and end of the extended emission region, a high \sigmagas\ nuclear region ($\sim240$ \kms), and a low \sigmagas\ region in the cone ($\sim120$ \kms).

Our SITELLE \halpha\ observations, in combination with radio interferometry
observations \citep{Augusto2006a, Savini2019} and Chandra X-ray imaging \citep{Sonkamble2015}, point to at least three epochs of AGN outflow activity.
A very young milli-arcsecond scale North-South jet system is identified in \citet{Augusto2006a}.
Two large-scale ($\sim$600 kpc) fading radio lobes to the East-West in Figure \ref{fig:precession} are remnants of an older episode of activity as discussed in \citet{Savini2019}.
There is a region of flat spectral index radio sources that is located at a similar orientation but at a much greater extent to the \halpha\ cone, and marks another episode of AGN activity.
X-ray cavities associated with each of these episodes are observed in \citet{Sonkamble2015}.
We propose that these data point to three episodes of AGN activity, with the \halpha\ cone associated with an intermediate-age epoch of AGN activity.

Further information is needed to constrain the source of the ionizing radiation in the cone of Obj1757C.
The source of ionization will provide us information on the physical processes of kinetic mode AGN in galaxy clusters, and indicate how metal-rich gases are spread through the ICM from the cluster BCG.
High spatial resolution \halpha\ imaging or IFU data and high-resolution IR imaging from JWST, could provide information on the structure of the cone and clumpy regions seen in the HST imaging, and possible star formation or shock activity causing the ionization of the gas.
It would also be useful for determining the smaller-scale structure of the \halpha\ cone, providing information on whether the cone is moving outward from the BCG or is a filament falling back onto the BCG.
Additionally, deeper, higher resolution radio interferometry of the flat-spectrum radio sources in A2390 would be useful for determining the properties of the intermediate-age episode of AGN activity.

\section*{Acknowledgements}

We thank the anonymous referee for their assistance with this article.
The research of LYA is supported by an NSERC Discovery grant to HKCY.
JHL thanks the NSERC Discovery grant program, the Discovery Accelerator Supplements program and the Canada Research Chair program.
Based on observations obtained at the Canada-France-Hawaii Telescope (CFHT) which is operated from the summit of Maunakea by the National Research Council of Canada, the Institut National des Sciences de l'Univers of the Centre National de la Recherche Scientifique of France, and the University of Hawaii. The observations at the Canada-France-Hawaii Telescope were performed with care and respect from the summit of Maunakea which is a significant cultural and historic site, and we thank the indigenous and resident Hawaiian population for being our gracious hosts.
Based on observations obtained with SITELLE, a joint project between Université Laval, ABB-Bomem, Université de Montréal and the CFHT with funding support from the Canada Foundation for Innovation (CFI), the National Sciences and Engineering Research Council of Canada (NSERC), Fond de Recheche du Québec - Nature et Technologies (FRQNT) and CFHT.
This research made use of Photutils, an Astropy package for detection and photometry of astronomical sources (Bradley et al. 2020).

\section*{Data Availability}

The data underlying this article are available on the Canadian Astronomical Data Centre website, at https://www.cadc-ccda.hia-iha.nrc-cnrc.gc.ca/en/. The datasets can be downloaded using the proposal IDs listed in Table \ref{table:observations}.



\bibliographystyle{mnras}
\bibliography{main} 

\begin{thebibliography}{}
\makeatletter
\relax
\def\mn@urlcharsother{\let\do\@makeother \do\$\do\&\do\#\do\^\do\_\do\%\do\~}
\def\mn@doi{\begingroup\mn@urlcharsother \@ifnextchar [ {\mn@doi@}
  {\mn@doi@[]}}
\def\mn@doi@[#1]#2{\def\@tempa{#1}\ifx\@tempa\@empty \href
  {http://dx.doi.org/#2} {doi:#2}\else \href {http://dx.doi.org/#2} {#1}\fi
  \endgroup}
\def\mn@eprint#1#2{\mn@eprint@#1:#2::\@nil}
\def\mn@eprint@arXiv#1{\href {http://arxiv.org/abs/#1} {{\tt arXiv:#1}}}
\def\mn@eprint@dblp#1{\href {http://dblp.uni-trier.de/rec/bibtex/#1.xml}
  {dblp:#1}}
\def\mn@eprint@#1:#2:#3:#4\@nil{\def\@tempa {#1}\def\@tempb {#2}\def\@tempc
  {#3}\ifx \@tempc \@empty \let \@tempc \@tempb \let \@tempb \@tempa \fi \ifx
  \@tempb \@empty \def\@tempb {arXiv}\fi \@ifundefined
  {mn@eprint@\@tempb}{\@tempb:\@tempc}{\expandafter \expandafter \csname
  mn@eprint@\@tempb\endcsname \expandafter{\@tempc}}}

\bibitem[\protect\citeauthoryear{Abraham et~al.,}{Abraham
  et~al.}{1996}]{Abraham1996}
Abraham R.~G.,  et~al., 1996, \mn@doi [The Astrophysical Journal]
  {10.1086/177999}, 471, 694

\bibitem[\protect\citeauthoryear{Aharonia et~al.,}{Aharonia
  et~al.}{2018}]{Aharonia2018}
Aharonia F.,  et~al., 2018, \mn@doi [Publications of the Astronomical Society
  of Japan] {10.1093/pasj/psx138}, 70, 1

\bibitem[\protect\citeauthoryear{Allen, Ettori  \& Fabian}{Allen
  et~al.}{2001}]{Allen2001}
Allen S.~W.,  Ettori S.,   Fabian A.~C.,  2001, \mn@doi [Monthly Notices of the
  Royal Astronomical Society] {10.1046/j.1365-8711.2001.04318.x}, 324, 877

\bibitem[\protect\citeauthoryear{Allen, Schmidt  \& Fabian}{Allen
  et~al.}{2002}]{Allen2002}
Allen S.~W.,  Schmidt R.~W.,   Fabian A.~C.,  2002, \mn@doi [Monthly Notices of
  the Royal Astronomical Society] {10.1046/j.1365-8711.2002.05601.x}, 334, L11

\bibitem[\protect\citeauthoryear{Augusto, Edge  \& Chandler}{Augusto
  et~al.}{2006}]{Augusto2006a}
Augusto P.,  Edge A.~C.,   Chandler C.~J.,  2006, \mn@doi [Monthly Notices of
  the Royal Astronomical Society] {10.1111/j.1365-2966.2005.09965.x}, 367, 366

\bibitem[\protect\citeauthoryear{Baldwin, Phillips  \& Terlevich}{Baldwin
  et~al.}{1981}]{Baldwin1981}
Baldwin A.,  Phillips M.~M.,   Terlevich R.,  1981, \mn@doi [Publications of
  the Astronomical Society of the Pacific] {10.1086/130930}, 93, 817

\bibitem[\protect\citeauthoryear{Balogh \& Morris}{Balogh \&
  Morris}{2000}]{Balogh2000}
Balogh M.~L.,  Morris S.~L.,  2000, \mn@doi [Monthly Notices of the Royal
  Astronomical Society] {10.1046/j.1365-8711.2000.03826.x}, 318, 703

\bibitem[\protect\citeauthoryear{Blanton, Clarke, Sarazin, Randall  \&
  McNamara}{Blanton et~al.}{2010}]{Blanton2010}
Blanton E.~L.,  Clarke T.~E.,  Sarazin C.~L.,  Randall S.~W.,   McNamara B.~R.,
   2010, \mn@doi [Proceedings of the National Academy of Sciences of the United
  States of America] {10.1073/pnas.0913904107}, 107, 7174

\bibitem[\protect\citeauthoryear{Bradley et~al.,}{Bradley
  et~al.}{2020}]{larry_bradley_2020_4044744}
Bradley L.,  et~al., 2020, astropy/photutils: 1.0.0,
  \mn@doi{10.5281/zenodo.4044744}, \url
  {https://doi.org/10.5281/zenodo.4044744}

\bibitem[\protect\citeauthoryear{Brough, Tran, Sharp, von~der Linden  \&
  Couch}{Brough et~al.}{2011}]{Brough2011}
Brough S.,  Tran K.~V.,  Sharp R.~G.,  von~der Linden A.,   Couch W.~J.,  2011,
  \mn@doi [Monthly Notices of the Royal Astronomical Society: Letters]
  {10.1111/j.1745-3933.2011.01060.x}, 414, 80

\bibitem[\protect\citeauthoryear{Bézecourt \& Soucail}{Bézecourt \&
  Soucail}{1997}]{Bezecourt1997}
Bézecourt J.,  Soucail G.,  1997, Astronomy and Astrophysics, 317, 661

\bibitem[\protect\citeauthoryear{{Calzadilla} et~al.,}{{Calzadilla}
  et~al.}{2022}]{Calzadilla2022}
{Calzadilla} M.~S.,  et~al., 2022, \mn@doi [\apj] {10.3847/1538-4357/ac9790},
  \href {https://ui.adsabs.harvard.edu/abs/2022ApJ...940..140C} {940, 140}

\bibitem[\protect\citeauthoryear{Carlberg, Yee, Ellingson, Abraham, Gravel,
  Morris  \& Pritchet}{Carlberg et~al.}{1996}]{carlberg_galaxy_1996}
Carlberg Yee Ellingson Abraham Gravel Morris  Pritchet 1996, \mn@doi [ApJ]
  {10.1086/177125}, 462, 32

\bibitem[\protect\citeauthoryear{{Cavagnolo}, {Donahue}, {Voit}  \&
  {Sun}}{{Cavagnolo} et~al.}{2008}]{Cavagnolo2008}
{Cavagnolo} K.~W.,  {Donahue} M.,  {Voit} G.~M.,   {Sun} M.,  2008, \mn@doi
  [\apjl] {10.1086/591665}, \href
  {https://ui.adsabs.harvard.edu/abs/2008ApJ...683L.107C} {683, L107}

\bibitem[\protect\citeauthoryear{Cavagnolo, McNamara, Nulsen, Carilli, Jones
  \& Bîrzan}{Cavagnolo et~al.}{2010}]{Cavagnolo2010}
Cavagnolo K.~W.,  McNamara B.~R.,  Nulsen P.~E.,  Carilli C.~L.,  Jones C.,
  Bîrzan L.,  2010, \mn@doi [Astrophysical Journal]
  {10.1088/0004-637X/720/2/1066}, 720, 1066

\bibitem[\protect\citeauthoryear{Chen, Heinz  \& Enßlin}{Chen
  et~al.}{2019}]{Chen2019b}
Chen Y.~H.,  Heinz S.,   Enßlin T.~A.,  2019, \mn@doi [Monthly Notices of the
  Royal Astronomical Society] {10.1093/mnras/stz2256}, 489, 1939

\bibitem[\protect\citeauthoryear{Churazov, Forman, Jones  \&
  Böhringer}{Churazov et~al.}{2000}]{Churazov2000}
Churazov E.,  Forman W.,  Jones C.,   Böhringer H.,  2000, Astronomy and
  Astrophysics, 356, 788

\bibitem[\protect\citeauthoryear{{Churazov}, {Br{\"u}ggen}, {Kaiser},
  {B{\"o}hringer}  \& {Forman}}{{Churazov} et~al.}{2001}]{Churazov2001}
{Churazov} E.,  {Br{\"u}ggen} M.,  {Kaiser} C.~R.,  {B{\"o}hringer} H.,
  {Forman} W.,  2001, \mn@doi [\apj] {10.1086/321357}, \href
  {https://ui.adsabs.harvard.edu/abs/2001ApJ...554..261C} {554, 261}

\bibitem[\protect\citeauthoryear{{Churazov}, {Sazonov}, {Sunyaev}, {Forman},
  {Jones}  \& {B{\"o}hringer}}{{Churazov} et~al.}{2005}]{Churazov2005}
{Churazov} E.,  {Sazonov} S.,  {Sunyaev} R.,  {Forman} W.,  {Jones} C.,
  {B{\"o}hringer} H.,  2005, \mn@doi [\mnras]
  {10.1111/j.1745-3933.2005.00093.x}, \href
  {https://ui.adsabs.harvard.edu/abs/2005MNRAS.363L..91C} {363, L91}

\bibitem[\protect\citeauthoryear{Cid~Fernandes, Stasińska, Schlickmann,
  Mateus, Vale~Asari, Schoenell  \& Sodré}{Cid~Fernandes
  et~al.}{2010}]{CidFernandes2010}
Cid~Fernandes R.,  Stasińska G.,  Schlickmann M.~S.,  Mateus A.,  Vale~Asari
  N.,  Schoenell W.,   Sodré L.,  2010, \mn@doi [Monthly Notices of the Royal
  Astronomical Society] {10.1111/j.1365-2966.2009.16185.x}, 403, 1036

\bibitem[\protect\citeauthoryear{Ciocan, Ziegler, Verdugo, Papaderos, Fogarty,
  Donahue  \& Postman}{Ciocan et~al.}{2021}]{Ciocan2021}
Ciocan B.~I.,  Ziegler B.~L.,  Verdugo M.,  Papaderos P.,  Fogarty K.,  Donahue
  M.,   Postman M.,  2021, pp 1--31

\bibitem[\protect\citeauthoryear{Crawford, Allen, Ebeling, Edge  \&
  Fabian}{Crawford et~al.}{1999}]{Crawford1999}
Crawford C.~S.,  Allen S.~W.,  Ebeling H.,  Edge A.~C.,   Fabian A.~C.,  1999,
  \mn@doi [Monthly Notices of the Royal Astronomical Society]
  {10.1046/j.1365-8711.1999.02583.x}, 306, 857

\bibitem[\protect\citeauthoryear{David et~al.,}{David et~al.}{2011}]{David2011}
David L.~P.,  et~al., 2011, \mn@doi [Astrophysical Journal]
  {10.1088/0004-637X/728/2/162}, 728

\bibitem[\protect\citeauthoryear{{Doj{\v{c}}inovi{\'c}},
  {Kova{\v{c}}evi{\'c}-Doj{\v{c}}inovi{\'c}}  \&
  {Popovi{\'c}}}{{Doj{\v{c}}inovi{\'c}} et~al.}{2022}]{Dojcinovic2022}
{Doj{\v{c}}inovi{\'c}} I.,  {Kova{\v{c}}evi{\'c}-Doj{\v{c}}inovi{\'c}} J.,
  {Popovi{\'c}} L.~{\v{C}}.,  2022, arXiv e-prints, \href
  {https://ui.adsabs.harvard.edu/abs/2022arXiv220410036D} {p. arXiv:2204.10036}

\bibitem[\protect\citeauthoryear{Drissen et~al.,}{Drissen
  et~al.}{2019}]{Drissen2019}
Drissen L.,  et~al., 2019, \mn@doi [Monthly Notices of the Royal Astronomical
  Society] {10.1093/mnras/stz627}, 485, 3930

\bibitem[\protect\citeauthoryear{{Duan} \& {Guo}}{{Duan} \&
  {Guo}}{2018}]{Duan2018}
{Duan} X.,  {Guo} F.,  2018, \mn@doi [\apj] {10.3847/1538-4357/aac9ba}, \href
  {https://ui.adsabs.harvard.edu/abs/2018ApJ...861..106D} {861, 106}

\bibitem[\protect\citeauthoryear{{Durret}, {Wakamatsu}, {Nagayama}, {Adami}  \&
  {Biviano}}{{Durret} et~al.}{2015}]{Durret2015}
{Durret} F.,  {Wakamatsu} K.,  {Nagayama} T.,  {Adami} C.,   {Biviano} A.,
  2015, \mn@doi [\aap] {10.1051/0004-6361/201526531}, \href
  {https://ui.adsabs.harvard.edu/abs/2015A&A...583A.124D} {583, A124}

\bibitem[\protect\citeauthoryear{{Dursi} \& {Pfrommer}}{{Dursi} \&
  {Pfrommer}}{2008}]{Dursi2008}
{Dursi} L.~J.,  {Pfrommer} C.,  2008, \mn@doi [\apj] {10.1086/529371}, \href
  {https://ui.adsabs.harvard.edu/abs/2008ApJ...677..993D} {677, 993}

\bibitem[\protect\citeauthoryear{Ea et~al.,}{Ea et~al.}{2010}]{Ea2010a}
Ea C. h. P. O.~D.,  et~al., 2010, The Astrophysical Journal

\bibitem[\protect\citeauthoryear{Egami et~al.,}{Egami et~al.}{2006}]{Egami2006}
Egami E.,  et~al., 2006, \mn@doi [The Astrophysical Journal] {10.1086/504519},
  647, 922

\bibitem[\protect\citeauthoryear{{Fabian}}{{Fabian}}{1994}]{Fabian1994}
{Fabian} A.~C.,  1994, \mn@doi [\araa] {10.1146/annurev.aa.32.090194.001425},
  \href {https://ui.adsabs.harvard.edu/abs/1994ARA&A..32..277F} {32, 277}

\bibitem[\protect\citeauthoryear{{Fabian}}{{Fabian}}{2012}]{Fabian2012}
{Fabian} A.~C.,  2012, \mn@doi [\araa] {10.1146/annurev-astro-081811-125521},
  \href {https://ui.adsabs.harvard.edu/abs/2012ARA&A..50..455F} {50, 455}

\bibitem[\protect\citeauthoryear{Fabian, Sanders, Crawford, Conselice,
  Gallagher  \& Wyse}{Fabian et~al.}{2003}]{Fabian2003}
Fabian A.~C.,  Sanders J.~S.,  Crawford C.~S.,  Conselice C.~J.,  Gallagher
  J.~S.,   Wyse R.~F.,  2003, \mn@doi [Monthly Notices of the Royal
  Astronomical Society] {10.1046/j.1365-8711.2003.06856.x}, 344, 48

\bibitem[\protect\citeauthoryear{Fabian, Sanders, Taylor  \& Allen}{Fabian
  et~al.}{2005}]{Fabian2005}
Fabian A.~C.,  Sanders J.~S.,  Taylor G.~B.,   Allen S.~W.,  2005, \mn@doi
  [Monthly Notices of the Royal Astronomical Society: Letters]
  {10.1111/j.1745-3933.2005.00037.x}, 360

\bibitem[\protect\citeauthoryear{Fabian, Sanders, Williams, Lazarian, Ferland
  \& Johnstone}{Fabian et~al.}{2011}]{Fabian2011}
Fabian A.~C.,  Sanders J.~S.,  Williams R. J.~R.,  Lazarian A.,  Ferland G.~J.,
    Johnstone R.~M.,  2011, \mn@doi [Monthly Notices of the Royal Astronomical
  Society] {10.1111/j.1365-2966.2011.19034.x}, 417, 172

\bibitem[\protect\citeauthoryear{Fabian et~al.,}{Fabian
  et~al.}{2016}]{Fabian2016}
Fabian A.~C.,  et~al., 2016, \mn@doi [Monthly Notices of the Royal Astronomical
  Society] {10.1093/mnras/stw1350}, 461, 922

\bibitem[\protect\citeauthoryear{{Fabian}, {ZuHone}  \& {Walker}}{{Fabian}
  et~al.}{2022}]{Fabian2021}
{Fabian} A.~C.,  {ZuHone} J.~A.,   {Walker} S.~A.,  2022, \mn@doi [\mnras]
  {10.1093/mnras/stab3655}, \href
  {https://ui.adsabs.harvard.edu/abs/2022MNRAS.510.4000F} {510, 4000}

\bibitem[\protect\citeauthoryear{Fisher, Illingworth  \& Franx}{Fisher
  et~al.}{1995}]{Fisher1995}
Fisher D.,  Illingworth G.,   Franx M.,  1995, \mn@doi [The Astrophysical
  Journal] {10.1086/175100}, 438, 539

\bibitem[\protect\citeauthoryear{Fogarty, Postman, Connor, Donahue  \&
  Moustakas}{Fogarty et~al.}{2015}]{Fogarty2015}
Fogarty K.,  Postman M.,  Connor T.,  Donahue M.,   Moustakas J.,  2015,
  \mn@doi [Astrophysical Journal] {10.1088/0004-637X/813/2/117}, 813, 117

\bibitem[\protect\citeauthoryear{{Fort}}{{Fort}}{1994}]{Fort1994}
{Fort} B.,  1994, {Gravitational Lensing Studies with Hst: Dark Matter and High
  Redshift Galaxies -- CYCLE4 High}, HST Proposal ID 5352. Cycle 4

\bibitem[\protect\citeauthoryear{Frye \& Broadhurst}{Frye \&
  Broadhurst}{1998}]{Frye1998}
Frye B.,  Broadhurst T.,  1998, \mn@doi [The Astrophysical Journal]
  {10.1086/311361}, 499, L115

\bibitem[\protect\citeauthoryear{{Gaspari}, {Ruszkowski}  \&
  {Sharma}}{{Gaspari} et~al.}{2012}]{Gaspari2012}
{Gaspari} M.,  {Ruszkowski} M.,   {Sharma} P.,  2012, \mn@doi [\apj]
  {10.1088/0004-637X/746/1/94}, \href
  {https://ui.adsabs.harvard.edu/abs/2012ApJ...746...94G} {746, 94}

\bibitem[\protect\citeauthoryear{{Giacintucci}, {Markevitch},
  {Johnston-Hollitt}, {Wik}, {Wang}  \& {Clarke}}{{Giacintucci}
  et~al.}{2020}]{Giacintucci2020}
{Giacintucci} S.,  {Markevitch} M.,  {Johnston-Hollitt} M.,  {Wik} D.~R.,
  {Wang} Q.~H.~S.,   {Clarke} T.~E.,  2020, \mn@doi [\apj]
  {10.3847/1538-4357/ab6a9d}, \href
  {https://ui.adsabs.harvard.edu/abs/2020ApJ...891....1G} {891, 1}

\bibitem[\protect\citeauthoryear{Haines et~al.,}{Haines
  et~al.}{2013}]{Haines2013}
Haines C.~P.,  et~al., 2013, \mn@doi [Astrophysical Journal]
  {10.1088/0004-637X/775/2/126}, 775

\bibitem[\protect\citeauthoryear{Hamer et~al.,}{Hamer et~al.}{2016}]{Hamer2016}
Hamer S.~L.,  et~al., 2016, \mn@doi [Monthly Notices of the Royal Astronomical
  Society] {10.1093/mnras/stw1054}, 460, 1758

\bibitem[\protect\citeauthoryear{{Heckman}, {Baum}, {van Breugel}  \&
  {McCarthy}}{{Heckman} et~al.}{1989}]{Heckman1989}
{Heckman} T.~M.,  {Baum} S.~A.,  {van Breugel} W.~J.~M.,   {McCarthy} P.,
  1989, \mn@doi [\apj] {10.1086/167181}, \href
  {https://ui.adsabs.harvard.edu/abs/1989ApJ...338...48H} {338, 48}

\bibitem[\protect\citeauthoryear{Hlavacek-Larrondo \& Fabian}{Hlavacek-Larrondo
  \& Fabian}{2011}]{Hlavacek-Larrondo2011}
Hlavacek-Larrondo J.,  Fabian A.~C.,  2011, \mn@doi [Monthly Notices of the
  Royal Astronomical Society] {10.1111/j.1365-2966.2010.18138.x}, 413, 313

\bibitem[\protect\citeauthoryear{{Hlavacek-Larrondo}, {Fabian}, {Edge},
  {Ebeling}, {Allen}, {Sanders}  \& {Taylor}}{{Hlavacek-Larrondo}
  et~al.}{2013}]{Hlavacek-Larrondo2013}
{Hlavacek-Larrondo} J.,  {Fabian} A.~C.,  {Edge} A.~C.,  {Ebeling} H.,  {Allen}
  S.~W.,  {Sanders} J.~S.,   {Taylor} G.~B.,  2013, \mn@doi [\mnras]
  {10.1093/mnras/stt283}, \href
  {https://ui.adsabs.harvard.edu/abs/2013MNRAS.431.1638H} {431, 1638}

\bibitem[\protect\citeauthoryear{{Hogan} et~al.,}{{Hogan}
  et~al.}{2017}]{Hogan2017}
{Hogan} M.~T.,  et~al., 2017, \mn@doi [\apj] {10.3847/1538-4357/aa9af3}, \href
  {https://ui.adsabs.harvard.edu/abs/2017ApJ...851...66H} {851, 66}

\bibitem[\protect\citeauthoryear{Hudson, Mittal, Reiprich, Nulsen, Andernach
  \& Sarazin}{Hudson et~al.}{2010}]{Hudson2010}
Hudson D.~S.,  Mittal R.,  Reiprich T.~H.,  Nulsen P.~E.,  Andernach H.,
  Sarazin C.~L.,  2010, \mn@doi [Astronomy and Astrophysics]
  {10.1051/0004-6361/200912377}, 513, 1

\bibitem[\protect\citeauthoryear{Hutchings \& Balogh}{Hutchings \&
  Balogh}{2000}]{Hutchings2000}
Hutchings J.~B.,  Balogh M.~L.,  2000, \mn@doi [The Astronomical Journal]
  {10.1086/301253}, 119, 1123

\bibitem[\protect\citeauthoryear{{Jimmy}, Tran, Brough, Gebhardt, Von
  Der~Linden, Couch  \& Sharp}{{Jimmy} et~al.}{2013}]{Jimmy2013}
{Jimmy} Tran K.~V.,  Brough S.,  Gebhardt K.,  Von Der~Linden A.,  Couch W.~J.,
    Sharp R.,  2013, \mn@doi [Astrophysical Journal]
  {10.1088/0004-637X/778/2/171}, 778

\bibitem[\protect\citeauthoryear{Kauffmann et~al.,}{Kauffmann
  et~al.}{2003}]{Kauffmann2003}
Kauffmann G.,  et~al., 2003, \mn@doi [Monthly Notices of the Royal Astronomical
  Society] {10.1111/j.1365-2966.2003.07154.x}, 346, 1055

\bibitem[\protect\citeauthoryear{Kennicutt}{Kennicutt}{1997}]{Kennicutt1998}
Kennicutt R.~C.,  1997, \mn@doi [The Astrophysical Journal] {10.1086/305588},
  498, 541

\bibitem[\protect\citeauthoryear{Kewley, Heisler, Dopita  \& Lumsden}{Kewley
  et~al.}{2001}]{Kewley2001}
Kewley L.~J.,  Heisler C.~A.,  Dopita M.~A.,   Lumsden S.,  2001, \mn@doi [The
  Astrophysical Journal Supplement Series] {10.1086/318944}, 132, 37

\bibitem[\protect\citeauthoryear{Lauer, Postman, Strauss, Graves  \&
  Chisari}{Lauer et~al.}{2014}]{Lauer2014}
Lauer T.~R.,  Postman M.,  Strauss M.~A.,  Graves G.~J.,   Chisari N.~E.,
  2014, \mn@doi [Astrophysical Journal] {10.1088/0004-637X/797/2/82}, 797

\bibitem[\protect\citeauthoryear{Li}{Li}{2020}]{Li2020}
Li A.,  2020, \mn@doi [Nature Astronomy] {10.1038/s41550-020-1051-1}, 4, 339

\bibitem[\protect\citeauthoryear{{Li} \& {Bryan}}{{Li} \&
  {Bryan}}{2014a}]{Li2014b}
{Li} Y.,  {Bryan} G.~L.,  2014a, \mn@doi [\apj] {10.1088/0004-637X/789/1/54},
  \href {https://ui.adsabs.harvard.edu/abs/2014ApJ...789...54L} {789, 54}

\bibitem[\protect\citeauthoryear{{Li} \& {Bryan}}{{Li} \&
  {Bryan}}{2014b}]{Li2014}
{Li} Y.,  {Bryan} G.~L.,  2014b, \mn@doi [\apj] {10.1088/0004-637X/789/2/153},
  \href {https://ui.adsabs.harvard.edu/abs/2014ApJ...789..153L} {789, 153}

\bibitem[\protect\citeauthoryear{Li, Yee  \& Ellingson}{Li
  et~al.}{2009}]{Li2009}
Li I.~H.,  Yee H.~K.,   Ellingson E.,  2009, \mn@doi [Astrophysical Journal]
  {10.1088/0004-637X/698/1/83}, 698, 83

\bibitem[\protect\citeauthoryear{{Lin} \& {Mohr}}{{Lin} \&
  {Mohr}}{2004}]{Lin2004}
{Lin} Y.-T.,  {Mohr} J.~J.,  2004, \mn@doi [\apj] {10.1086/425412}, \href
  {https://ui.adsabs.harvard.edu/abs/2004ApJ...617..879L} {617, 879}

\bibitem[\protect\citeauthoryear{{Lin} \& {Mohr}}{{Lin} \&
  {Mohr}}{2007}]{Lin2007}
{Lin} Y.-T.,  {Mohr} J.~J.,  2007, \mn@doi [\apjs] {10.1086/513565}, \href
  {https://ui.adsabs.harvard.edu/abs/2007ApJS..170...71L} {170, 71}

\bibitem[\protect\citeauthoryear{{Lin}, {Ostriker}  \& {Miller}}{{Lin}
  et~al.}{2010}]{lin_new_2010}
{Lin} Y.-T.,  {Ostriker} J.~P.,   {Miller} C.~J.,  2010, \mn@doi [\apj]
  {10.1088/0004-637X/715/2/1486}, \href
  {https://ui.adsabs.harvard.edu/abs/2010ApJ...715.1486L} {715, 1486}

\bibitem[\protect\citeauthoryear{{Liu} et~al.,}{{Liu} et~al.}{2021}]{Liu2021}
{Liu} Q.,  et~al., 2021, \mn@doi [\apj] {10.3847/1538-4357/abd71e}, \href
  {https://ui.adsabs.harvard.edu/abs/2021ApJ...908..228L} {908, 228}

\bibitem[\protect\citeauthoryear{Maier, Haines  \& Ziegler}{Maier
  et~al.}{2022}]{maier_star-formation_2022}
Maier C.,  Haines C.~P.,   Ziegler B.~L.,  2022, \mn@doi [A\&A]
  {10.1051/0004-6361/202141498}, 658, A190

\bibitem[\protect\citeauthoryear{Martin, Prunet  \& Drissen}{Martin
  et~al.}{2016}]{Martin2016}
Martin T.~B.,  Prunet S.,   Drissen L.,  2016, \mn@doi [Monthly Notices of the
  Royal Astronomical Society] {10.1093/mnras/stw2315}, 463, 4223

\bibitem[\protect\citeauthoryear{{McDonald}, {Veilleux}, {Rupke}  \&
  {Mushotzky}}{{McDonald} et~al.}{2010}]{McDonald2010}
{McDonald} M.,  {Veilleux} S.,  {Rupke} D. S.~N.,   {Mushotzky} R.,  2010,
  \mn@doi [\apj] {10.1088/0004-637X/721/2/1262}, \href
  {https://ui.adsabs.harvard.edu/abs/2010ApJ...721.1262M} {721, 1262}

\bibitem[\protect\citeauthoryear{McNamara \& Nulsen}{McNamara \&
  Nulsen}{2007}]{McNamara2007}
McNamara B.~R.,  Nulsen P.~E.,  2007, \mn@doi [Annual Review of Astronomy and
  Astrophysics] {10.1146/annurev.astro.45.051806.110625}, 45, 117

\bibitem[\protect\citeauthoryear{McNamara \& O'Connell}{McNamara \&
  O'Connell}{1993}]{McNamara1993}
McNamara B.~R.,  O'Connell R.~W.,  1993, \mn@doi [The Astronomical Journal]
  {10.1086/116440}, 105, 417

\bibitem[\protect\citeauthoryear{McNamara et~al.,}{McNamara
  et~al.}{2000}]{McNamara2000}
McNamara B.~R.,  et~al., 2000, \mn@doi [The Astrophysical Journal]
  {10.1086/312662}, 534, L135

\bibitem[\protect\citeauthoryear{McNamara et~al.,}{McNamara
  et~al.}{2006}]{McNamara2006}
McNamara B.~R.,  et~al., 2006, \mn@doi [The Astrophysical Journal]
  {10.1086/505859}, 648, 164

\bibitem[\protect\citeauthoryear{McNamara, Kazemzadeh, Rafferty, Birzan,
  Nulsen, Kirkpatrick  \& Wise}{McNamara
  et~al.}{2009}]{mcnamara_energetic_2009}
McNamara B.~R.,  Kazemzadeh F.,  Rafferty D.~A.,  Birzan L.,  Nulsen P. E.~J.,
  Kirkpatrick C.~C.,   Wise M.~W.,  2009, \mn@doi [ApJ]
  {10.1088/0004-637X/698/1/594}, 698, 594

\bibitem[\protect\citeauthoryear{McNamara, Russell, Nulsen, Hogan, Fabian,
  Pulido  \& Edge}{McNamara et~al.}{2016}]{McNamara2016}
McNamara B.~R.,  Russell H.~R.,  Nulsen P. E.~J.,  Hogan M.~T.,  Fabian A.~C.,
  Pulido F.,   Edge A.~C.,  2016, \mn@doi [The Astrophysical Journal]
  {10.3847/0004-637x/830/2/79}, 830, 79

\bibitem[\protect\citeauthoryear{{Montes}}{{Montes}}{2022}]{montes_faint_2022}
{Montes} M.,  2022, \mn@doi [Nature Astronomy] {10.1038/s41550-022-01616-z},
  \href {https://ui.adsabs.harvard.edu/abs/2022NatAs...6..308M} {6, 308}

\bibitem[\protect\citeauthoryear{Morisset, Delgado-Inglada  \&
  Flores-Fajardo}{Morisset et~al.}{2015}]{Morisset2015}
Morisset C.,  Delgado-Inglada G.,   Flores-Fajardo N.,  2015, Revista Mexicana
  de Astronomia y Astrofisica, 51, 101

\bibitem[\protect\citeauthoryear{Oegerle \& Hoessel}{Oegerle \&
  Hoessel}{1991}]{Oegerle1991}
Oegerle W.~R.,  Hoessel J.~G.,  1991, \mn@doi [The Astrophysical Journal]
  {10.1086/170165}, 375, 15

\bibitem[\protect\citeauthoryear{{Patnaik}, {Browne}, {Wilkinson}  \&
  {Wrobel}}{{Patnaik} et~al.}{1992}]{Patnaik1992}
{Patnaik} A.~R.,  {Browne} I. W.~A.,  {Wilkinson} P.~N.,   {Wrobel} J.~M.,
  1992, \mn@doi [\mnras] {10.1093/mnras/254.4.655}, \href
  {https://ui.adsabs.harvard.edu/abs/1992MNRAS.254..655P} {254, 655}

\bibitem[\protect\citeauthoryear{Pelló et~al.,}{Pelló
  et~al.}{1999}]{Pello1999}
Pelló R.,  et~al., 1999, Astronomy and Astrophysics, 346, 359

\bibitem[\protect\citeauthoryear{Pierre, Le~Borgne, Soucail  \& Kneib}{Pierre
  et~al.}{1996}]{Pierre1996}
Pierre M.,  Le~Borgne J.~F.,  Soucail G.,   Kneib J.~P.,  1996, Astronomy and
  Astrophysics, 311, 413

\bibitem[\protect\citeauthoryear{{Pope}, {Babul}, {Pavlovski}, {Bower}  \&
  {Dotter}}{{Pope} et~al.}{2010}]{Pope2010}
{Pope} E. C.~D.,  {Babul} A.,  {Pavlovski} G.,  {Bower} R.~G.,   {Dotter} A.,
  2010, \mn@doi [\mnras] {10.1111/j.1365-2966.2010.16816.x}, \href
  {https://ui.adsabs.harvard.edu/abs/2010MNRAS.406.2023P} {406, 2023}

\bibitem[\protect\citeauthoryear{{Revaz}, {Combes}  \& {Salom{\'e}}}{{Revaz}
  et~al.}{2008}]{Revaz2008}
{Revaz} Y.,  {Combes} F.,   {Salom{\'e}} P.,  2008, \mn@doi [\aap]
  {10.1051/0004-6361:20078915}, \href
  {https://ui.adsabs.harvard.edu/abs/2008A&A...477L..33R} {477, L33}

\bibitem[\protect\citeauthoryear{Rhea, Rousseau-Nepton, Covington, Alcorn,
  Vigneron, Hlavacek-Larrondo  \& Guité}{Rhea et~al.}{2021a}]{RheaLUCI2021}
Rhea C.~L.,  Rousseau-Nepton L.,  Covington J.,  Alcorn L.,  Vigneron B.,
  Hlavacek-Larrondo J.,   Guité L.-S.,  2021a, crhea93/LUCI: Luci Updates,
  \mn@doi{10.5281/zenodo.5730149}, \url
  {https://doi.org/10.5281/zenodo.5730149}

\bibitem[\protect\citeauthoryear{Rhea, Hlavacek-Larrondo, Rousseau-Nepton  \&
  Prunet}{Rhea et~al.}{2021b}]{Rhea2021a}
Rhea C.,  Hlavacek-Larrondo J.,  Rousseau-Nepton L.,   Prunet S.,  2021b,
  \mn@doi [Research Notes of the AAS] {10.3847/2515-5172/ac3dfe}, 5, 276

\bibitem[\protect\citeauthoryear{Rhea, Rousseau-Nepton, Prunet, Prasow-Emond,
  Hlavacek-Larrondo, Asari, Grasha  \& Perreault-Levasseur}{Rhea
  et~al.}{2021c}]{Rhea2021}
Rhea C.,  Rousseau-Nepton L.,  Prunet S.,  Prasow-Emond M.,  Hlavacek-Larrondo
  J.,  Asari N.~V.,  Grasha K.,   Perreault-Levasseur L.,  2021c, \mn@doi [The
  Astrophysical Journal] {10.3847/1538-4357/abe627}, 910, 129

\bibitem[\protect\citeauthoryear{{Rousseau-Nepton} et~al.,}{{Rousseau-Nepton}
  et~al.}{2019}]{RousseauNepton2019}
{Rousseau-Nepton} L.,  et~al., 2019, \mn@doi [\mnras] {10.1093/mnras/stz2455},
  \href {https://ui.adsabs.harvard.edu/abs/2019MNRAS.489.5530R} {489, 5530}

\bibitem[\protect\citeauthoryear{{Russell}, {McNamara}, {Edge}, {Hogan}, {Main}
   \& {Vantyghem}}{{Russell} et~al.}{2013}]{Russell2013}
{Russell} H.~R.,  {McNamara} B.~R.,  {Edge} A.~C.,  {Hogan} M.~T.,  {Main}
  R.~A.,   {Vantyghem} A.~N.,  2013, \mn@doi [\mnras] {10.1093/mnras/stt490},
  \href {https://ui.adsabs.harvard.edu/abs/2013MNRAS.432..530R} {432, 530}

\bibitem[\protect\citeauthoryear{{Russell} et~al.,}{{Russell}
  et~al.}{2017}]{Russell2017}
{Russell} H.~R.,  et~al., 2017, \mn@doi [\mnras] {10.1093/mnras/stx2255}, \href
  {https://ui.adsabs.harvard.edu/abs/2017MNRAS.472.4024R} {472, 4024}

\bibitem[\protect\citeauthoryear{{Russell} et~al.,}{{Russell}
  et~al.}{2019}]{Russell2019}
{Russell} H.~R.,  et~al., 2019, \mn@doi [\mnras] {10.1093/mnras/stz2719}, \href
  {https://ui.adsabs.harvard.edu/abs/2019MNRAS.490.3025R} {490, 3025}

\bibitem[\protect\citeauthoryear{{Salom{\'e}} et~al.,}{{Salom{\'e}}
  et~al.}{2006}]{Salome2006}
{Salom{\'e}} P.,  et~al., 2006, \mn@doi [\aap] {10.1051/0004-6361:20054745},
  \href {https://ui.adsabs.harvard.edu/abs/2006A&A...454..437S} {454, 437}

\bibitem[\protect\citeauthoryear{Savini et~al.,}{Savini
  et~al.}{2018}]{Savini2019}
Savini F.,  et~al., 2018, \mn@doi [Astronomy and Astrophysics]
  {10.1051/0004-6361/201833882}, 622

\bibitem[\protect\citeauthoryear{{Sharma}, {McCourt}, {Quataert}  \&
  {Parrish}}{{Sharma} et~al.}{2012}]{Sharma2012}
{Sharma} P.,  {McCourt} M.,  {Quataert} E.,   {Parrish} I.~J.,  2012, \mn@doi
  [\mnras] {10.1111/j.1365-2966.2011.20246.x}, \href
  {https://ui.adsabs.harvard.edu/abs/2012MNRAS.420.3174S} {420, 3174}

\bibitem[\protect\citeauthoryear{Sivanandam, Rieke  \& Rieke}{Sivanandam
  et~al.}{2014}]{Sivanandam2014}
Sivanandam S.,  Rieke M.~J.,   Rieke G.~H.,  2014, \mn@doi [Astrophysical
  Journal] {10.1088/0004-637X/796/2/89}, 796

\bibitem[\protect\citeauthoryear{{Smail}, {Ivison}, {Blain}  \&
  {Kneib}}{{Smail} et~al.}{2002}]{Smail2002}
{Smail} I.,  {Ivison} R.~J.,  {Blain} A.~W.,   {Kneib} J.~P.,  2002, \mn@doi
  [\mnras] {10.1046/j.1365-8711.2002.05203.x}, \href
  {https://ui.adsabs.harvard.edu/abs/2002MNRAS.331..495S} {331, 495}

\bibitem[\protect\citeauthoryear{Smith \& Donohoe}{Smith \&
  Donohoe}{2021}]{Smith2021}
Smith M.~D.,  Donohoe J.,  2021, \mn@doi [Monthly Notices of the Royal
  Astronomical Society] {10.1093/mnras/stab044}, 502, 423

\bibitem[\protect\citeauthoryear{Sonkamble, Vagshette, Pawar  \&
  Patil}{Sonkamble et~al.}{2014}]{Sonkamble2015}
Sonkamble S.~S.,  Vagshette N.~D.,  Pawar P.~K.,   Patil M.~K.,  2014, \mn@doi
  [Astrophysics and Space Science] {10.1007/s10509-015-2508-z}, 359, 1

\bibitem[\protect\citeauthoryear{Stroe, Sobral, Paulino-Afonso, Alegre, Calhau,
  Santos  \& van Weeren}{Stroe et~al.}{2017}]{Stroe2017}
Stroe A.,  Sobral D.,  Paulino-Afonso A.,  Alegre L.,  Calhau J.,  Santos S.,
  van Weeren R.,  2017, \mn@doi [Monthly Notices of the Royal Astronomical
  Society] {10.1093/mnras/stw2939}, 465, 2916

\bibitem[\protect\citeauthoryear{Su, Nulsen, Kraft, Forman, Jones, Irwin,
  Randall  \& Churazov}{Su et~al.}{2017}]{Su2017}
Su Y.,  Nulsen P. E.~J.,  Kraft R.~P.,  Forman W.~R.,  Jones C.,  Irwin J.~A.,
  Randall S.~W.,   Churazov E.,  2017, \mn@doi [The Astrophysical Journal]
  {10.3847/1538-4357/aa8954}, 847, 94

\bibitem[\protect\citeauthoryear{Tran, Moustakas, Gonzalez, Bai, Zaritsky  \&
  Kautsch}{Tran et~al.}{2008}]{Tran2008}
Tran K.-V.~H.,  Moustakas J.,  Gonzalez A.~H.,  Bai L.,  Zaritsky D.,   Kautsch
  S.~J.,  2008, \mn@doi [The Astrophysical Journal] {10.1086/591422}, 683, L17

\bibitem[\protect\citeauthoryear{{Tremblay} et~al.,}{{Tremblay}
  et~al.}{2015}]{Tremblay2015}
{Tremblay} G.~R.,  et~al., 2015, \mn@doi [\mnras] {10.1093/mnras/stv1151},
  \href {https://ui.adsabs.harvard.edu/abs/2015MNRAS.451.3768T} {451, 3768}

\bibitem[\protect\citeauthoryear{{Tremblay} et~al.,}{{Tremblay}
  et~al.}{2018}]{Tremblay2018}
{Tremblay} G.~R.,  et~al., 2018, \mn@doi [\apj] {10.3847/1538-4357/aad6dd},
  \href {https://ui.adsabs.harvard.edu/abs/2018ApJ...865...13T} {865, 13}

\bibitem[\protect\citeauthoryear{Ueda et~al.,}{Ueda et~al.}{2018}]{Ueda2018}
Ueda S.,  et~al., 2018, \mn@doi [The Astrophysical Journal]
  {10.3847/1538-4357/aadd9d}, 866, 48

\bibitem[\protect\citeauthoryear{{Vantyghem}, {McNamara}, {Russell}, {Main},
  {Nulsen}, {Wise}, {Hoekstra}  \& {Gitti}}{{Vantyghem}
  et~al.}{2014}]{Vantyghem2014}
{Vantyghem} A.~N.,  {McNamara} B.~R.,  {Russell} H.~R.,  {Main} R.~A.,
  {Nulsen} P.~E.~J.,  {Wise} M.~W.,  {Hoekstra} H.,   {Gitti} M.,  2014,
  \mn@doi [\mnras] {10.1093/mnras/stu1030}, \href
  {https://ui.adsabs.harvard.edu/abs/2014MNRAS.442.3192V} {442, 3192}

\bibitem[\protect\citeauthoryear{{Voit}, {Meece}, {Li}, {O'Shea}, {Bryan}  \&
  {Donahue}}{{Voit} et~al.}{2017}]{Voit2017}
{Voit} G.~M.,  {Meece} G.,  {Li} Y.,  {O'Shea} B.~W.,  {Bryan} G.~L.,
  {Donahue} M.,  2017, \mn@doi [\apj] {10.3847/1538-4357/aa7d04}, \href
  {https://ui.adsabs.harvard.edu/abs/2017ApJ...845...80V} {845, 80}

\bibitem[\protect\citeauthoryear{Von Der~Linden, Best, Kauffmann  \& White}{Von
  Der~Linden et~al.}{2007}]{VonDerLinden2007}
Von Der~Linden A.,  Best P.~N.,  Kauffmann G.,   White S.~D.,  2007, \mn@doi
  [Monthly Notices of the Royal Astronomical Society]
  {10.1111/j.1365-2966.2007.11940.x}, 379, 867

\bibitem[\protect\citeauthoryear{{Wing} \& {Blanton}}{{Wing} \&
  {Blanton}}{2011}]{Wing2011}
{Wing} J.~D.,  {Blanton} E.~L.,  2011, \mn@doi [\aj]
  {10.1088/0004-6256/141/3/88}, \href
  {https://ui.adsabs.harvard.edu/abs/2011AJ....141...88W} {141, 88}

\bibitem[\protect\citeauthoryear{Yee, Ellingson, Abraham, Gravel, Carlberg,
  Smecker-Hane, Schade  \& Rigler}{Yee et~al.}{1996a}]{Yee1996a}
Yee H. K.~C.,  Ellingson E.,  Abraham R.~G.,  Gravel P.,  Carlberg R.~G.,
  Smecker-Hane T.~A.,  Schade D.,   Rigler M.,  1996a, \mn@doi [The
  Astrophysical Journal Supplement Series] {10.1086/192260}, 102, 289

\bibitem[\protect\citeauthoryear{Yee, Ellingson  \& Carlberg}{Yee
  et~al.}{1996b}]{Yee1996}
Yee H. K.~C.,  Ellingson E.,   Carlberg R.~G.,  1996b, \mn@doi [The
  Astrophysical Journal Supplement Series] {10.1086/192259}, 102, 269

\bibitem[\protect\citeauthoryear{{Zhang}, {Zhuravleva}, {Gendron-Marsolais},
  {Churazov}, {Schekochihin}  \& {Forman}}{{Zhang} et~al.}{2022}]{Zhang2022}
{Zhang} C.,  {Zhuravleva} I.,  {Gendron-Marsolais} M.-L.,  {Churazov} E.,
  {Schekochihin} A.~A.,   {Forman} W.~R.,  2022, \mn@doi [\mnras]
  {10.1093/mnras/stac2282}, \href
  {https://ui.adsabs.harvard.edu/abs/2022MNRAS.517..616Z} {517, 616}

\makeatother
\end{thebibliography}




\appendix


\bsp	
\label{lastpage}
\end{document}